\documentclass[%
 aip,
amsmath,amssymb,
reprint,
]{revtex4-1}

\usepackage{xcolor}
\usepackage{subcaption}
\usepackage{hyperref}
\usepackage{dirtytalk}
\usepackage{mathrsfs}

\usepackage{siunitx}
\DeclareSIUnit\angstrom{\text{Å}}

\usepackage{expl3}
\usepackage{calc}
\usepackage[version=4]{mhchem}

\usepackage{bm}

\usepackage[utf8]{inputenc}
\usepackage[T1]{fontenc}
\usepackage{mathptmx}
\usepackage{etoolbox}
\usepackage{afterpage}
\usepackage{placeins}

\usepackage{dcolumn}
\usepackage{booktabs}
\usepackage{multirow}
\usepackage{rotating}

\def\rA{\mathrm{A}}
\def\rD{\mathrm{D}}

\def\rIP{\mathrm{IP}}

\newcommand{\figshareAtoms}{
    Figure taken from E. M. Jahr and F. L. Hofmann (2026). "Screening Atom-Atom Systems for ICEC," Figshare.\cite{figshareAtoms} Author(s), licensed under a Creative Commons Attribution 4.0 License
}

\newcommand{\figshareMolecules}{
    Figure taken from E. M. Jahr and F. L. Hofmann (2026). "Screening Atom-Molecule Systems for ICEC," Figshare.\cite{figshareMolecules} Author(s), licensed under a Creative Commons Attribution 4.0 License
}

\makeatletter
\def\@email#1#2{%
 \endgroup
 \patchcmd{\titleblock@produce}
  {\frontmatter@RRAPformat}
  {\frontmatter@RRAPformat{\produce@RRAP{*#1\href{mailto:#2}{#2}}}\frontmatter@RRAPformat}
  {}{}
}%
\makeatother

\begin{document}

\preprint{AIP/123-QED}

\title{Where to Find ICEC? -- Screening 2442 Systems for Interparticle Coulombic Electron Capture}

\author{Fabian L. Hofmann}
    \thanks{These authors contributed equally to this work.}
    \affiliation{Institute of Physical and Theoretical Chemistry, University of Tübingen, Auf der Morgenstelle 18, 72076 Tübingen, Germany}
    \affiliation{Institute of Physical Chemistry, Karlsruhe Institute of Technology (KIT), Fritz-Haber-Weg 2, 76131 Karlsruhe, Germany}   

\author{Elena M. Jahr}
    \thanks{These authors contributed equally to this work.}
    \affiliation{Institute of Physical and Theoretical Chemistry, University of Tübingen, Auf der Morgenstelle 18, 72076 Tübingen, Germany}
    \affiliation{Center for Light-Matter Interaction, Sensors and Analytics (LISA+), University of Tübingen, Auf der Morgenstelle 15, 72076 Tübingen, Germany}

\author{Elke Fasshauer}
    \email{elke.fasshauer@pm.me}
    \affiliation{Institute of Physical and Theoretical Chemistry, University of Tübingen, Auf der Morgenstelle 18, 72076 Tübingen, Germany}
    \affiliation{Center for Light-Matter Interaction, Sensors and Analytics (LISA+), University of Tübingen, Auf der Morgenstelle 15, 72076 Tübingen, Germany}

\date{\today}

\begin{abstract}
Interparticle Coulombic Electron Capture (ICEC) provides an
environment-assisted pathway for electron attachment that can compete
with photorecombination. Here, we
present the first comprehensive survey of ICEC in atom-atom and atom-molecule systems,
screening 2442 combinations in search of promising candidates for future
experimental investigation.
We employ the efficient asymptotic approximation to
predict ICEC cross sections and electron
spectra. Intramolecular nuclear motion is included, while interparticle
nuclear dynamics is neglected to permit an extensive survey.
We identify several classes of atom–atom and atom–molecule systems with
favourable ICEC cross sections and ICEC-to-photorecombination ratios,
including halogen–halide systems and systems involving a proton and
diatomic molecules such as \ce{N2}, \ce{O2}, \ce{CO}, and \ce{NO}.
These findings point to atmospheric and astrochemical environments in which ICEC
may be relevant.
\end{abstract}

\pacs{}

\maketitle 

\section{Introduction}
Electron attachment processes are important in astro- and plasmaphysics\cite{tucker1978_astro, visentin2023_astro}, atmospheric chemistry\cite{Campbell2016_atmosphere}, electrochemistry~\cite{morgan1983_semiconductors}, and biochemistry~\cite{narayanan2023_dna}.
In 2009, an environment enabled electron attachment process was predicted theoretically:
the Interparticle Coulombic Electron Capture (ICEC).
\cite{gokhberg2009}
During this process, a free electron is captured by an electron
acceptor A and the excess energy is transferred to an electron donor D, which consequently gets ionized:
\begin{equation}
\label{eq:icec}
\mathrm{e}^-_k + \mathrm{A} + \mathrm{D} \to \mathrm{A}^- + \mathrm{D}^+ + \mathrm{e}^-_{k'}.
\end{equation}

ICEC has already been studied in systems such as noble gas dimers \cite{sisourat2018_icec}, proton water dimers\cite{molle2021_water}, and quantum dots\cite{pont2013_qd}.
Various theoretical methods and approximations have been developed to study the different ICEC pathways in these systems.
A comprehensive review on ICEC can be found in Ref.~\onlinecite{bande2023}.

In order for ICEC to occur, the process needs to be energetically accessible and sufficiently
efficient to compete with the photorecombination of the isolated electron acceptor as well as
with the two-center Dielectronic Recombination process (2CDR).
In 2CDR, the excess energy from electron attachment in the first unit excites the second unit, which consequently decays by emitting a photon rather than an electron \cite{mueller2010_2cdr}.
All these are electron scattering  or half-scattering processes and their efficiencies can thus be characterized by the scattering cross section. \cite{gokhberg2009,gokhberg2010,bande2023}

Currently, experiments are conducted and planned to verify ICEC.
In this context, the aim of this article is to guide these and future endeavours
by providing theoretical ICEC cross sections and spectra for a large variety of small systems.
To be precise, we study systems in different charge states consisting either of
two atoms from 23 elements, represented by 49 unique charged species, 
or of a single atom combined with one of six neutral diatomic molecules. 
These two classes yield 2174 and 268 combinations with at least one ICEC data point, respectively, resulting in a total of 2442 systems.


Considering that we attempt to scan the ICEC properties for many systems,
we desire a computationally efficient way to determine them. We therefore choose to use the
asymptotic approximation \cite{gokhberg2009,gokhberg2010,bande2023,jahr2026_intra}.
This approach has proven to yield qualitatively correct results for ICEC, as well as the Interparticle Coulombic Decay (ICD).
For ICD, its efficiency enabled
the studied system size to increase from clusters consisting of 13
to several
thousand atoms.\cite{Fasshauer13,Fasshauer16}
At finite interparticle distances, however,
the respective ICEC cross
sections represent
a lower bound to the true ICEC cross section, because terms involving
orbital overlap between the units are neglected\cite{Molle21a,senk2024_overlap}.

Recent studies have introduced the possibility to include the effect of both \emph{inter}particle\cite{jahr2025_inter} and \emph{intra}particle\cite{jahr2026_intra} nuclear dynamics
on the ICEC cross section within the asymptotic approximation.
The determination of ICEC cross sections including the impact of interparticle nuclear dynamics requires the relevant interparticle potentials and
therefore poses a significant additional effort for systems, where ICEC may not even be relevant.
For this reason, we ignore the impact of interparticle motion for the scan in this article.
On the other hand, taking the impact of intraparticle nuclear motion of the participating molecules into account only requires the vibrational energies of the ground and ionized states and the Franck-Condon factors between them\cite{jahr2026_intra}, which are accessible from the literature.
We therefore include the effect of intraparticle motion in our simulations.

%
The paper is organized as follows: 
Section~\ref{sec:theory} describes the theoretical approaches employed in this work, while  Section~\ref{sec:comp-details} outlines the computational details. 
In Section~\ref{sec:results}, we present and discuss our results, highlighting
recurring effects on the ICEC cross section and the most promising system candidates to observe ICEC.
Finally, we summarize our findings and their broader implications in Section~\ref{sec:conclusion}.

We use atomic units ($\hbar = m_e = e = 4\pi\epsilon_0 = 1$) throughout this paper, unless explicitly mentioned otherwise.
\section{Theory}
\label{sec:theory}

The ICEC process given in Eq.~\eqref{eq:icec} represents an inelastic electron scattering process \cite{gokhberg2009, taylor2006_scattering}.
Its efficiency is therefore characterized by the cross section
$\sigma(\varepsilon)$ of the process,
which gives the probability that an incoming electron with energy $\varepsilon$ is captured while an electron with energy $\varepsilon'$ is emitted\cite{bande2023, gokhberg2009}.

The asymptotic approximation to ICEC assumes A and D to be sufficiently separated for the electronic interaction between them to be weak.
First order perturbation theory then allows the cross section to be rationalized in terms of properties of the isolated units,\cite{gokhberg2009, bande2023}
\begin{equation}
    \label{eq:icec_electronic}
    \sigma(\varepsilon) 
    = 
    \frac{3 c^4 }{4 \pi} \,
    \frac{
    \sigma^\mathrm{PR}_\mathrm{A}(\varepsilon) \; 
    \sigma^\mathrm{PI}_\mathrm{D}(\omega)
    }{\omega^4 \, R^6}.
\end{equation}
Here, $\sigma^\mathrm{PR}_\mathrm{A}$ denotes the photorecombination (PR) cross section for an anionic electron acceptor A. For neutral A, the same quantity corresponds to radiative attachment or photoattachment.
Likewise, $\sigma^\mathrm{PI}_\mathrm{D}$ is the photoionization (PI) cross section of a neutral or cationic electron donor D, whereas for negatively charged D it corresponds to photodetachment.
$R$ is the distance between the center of masses of A and D and $c$ is the speed of light.

The transferred energy $\omega$ is obtained from energy conservation $\varepsilon + E_{\rA\rD} = \omega + E_{\rA^-\rD}$.
For infinite separation between A and D, this reduces to $\omega = \varepsilon + \rIP_{\rA^-}$, where $\rIP_{\rA^-}$ is the ionization potential (IP) of $\rA^-$.\cite{gokhberg2010}
For finite distances, however, the energy needs to be corrected for the attraction or repulsion between A and D.
Within the lowest order correction, the total energy is $E_\mathrm{AD} = E_\mathrm{A} + E_\mathrm{D} + \frac{Z_\rA Z_\rD}{R}$ and therefore
\begin{equation}
    \label{eq:omega}
    \omega = \varepsilon + \rIP_{\rA^-} + \frac{Z_\rD}{R},
\end{equation}
where $Z_\rD$ is the total charge of D.
When multiple neighbors are present, the energy can be transferred to any of them.
If all interactions between these neighbors can be neglected, the total ICEC cross section will be the sum of the individual contributions\cite{gokhberg2009, gokhberg2010}.

We analogously obtain the energy of the outgoing electron from the energy conservation, $\omega + E_\mathrm{\rA^-\rD} = \varepsilon' +  E_\mathrm{\rA^-\rD^+}$,
\begin{equation}
\label{eq:final_E_electronic}
\begin{aligned}
    \varepsilon' &= \omega - \mathrm{IP}_\mathrm{D} - \frac{(Z_\rA-1)}{R}\\
    &= \varepsilon + \mathrm{IP}_\mathrm{A^-} - \mathrm{IP}_\mathrm{D} + \frac{Z_\rD - Z_\rA +1 }{R},
\end{aligned}
\end{equation}
where $\rIP_\rD$ is the ionization potential of D and $Z_\rA$ is the charge of A.
If the transferred energy is insufficient to ionize D, then $\varepsilon'$ is negative, ICEC is energetically forbidden and the corresponding cross section is therefore zero.


Eq.~\eqref{eq:icec_electronic} is valid for atom-atom systems but can be applied to molecules when neglecting nuclear dynamics.
However, we recently extended the asymptotic approximation to include intramolecular nuclear degrees of freedom for both A and D demonstrating that nuclear dynamics changes the outgoing electron energy.\cite{jahr2026_intra}
We will apply this model with nuclear dynamics limited to D because we will only study systems where D is a molecule.

Within our Franck-Condon model for ICEC\cite{jahr2026_intra} we describe the initial and final vibronic states of D as a single term of the Born-Huang expansion \cite{ballhausen1972, born1996} and employ the Born–Oppenheimer approximation, neglecting rotational degrees of freedom.
Analogous to the derivation of Eq.~\eqref{eq:icec_electronic}, we assume A and D to be sufficiently separated and employ first order of perturbation theory for the interaction between A and D.
Averaging over the orientation of D removes the directional dependence due to a molecule not being spherically symmetric.
\cite{jahr2026_intra}

Beyond that, we employ the Condon approximation\cite{herzberg1950} for the vibronic transition of D.
This assumes that the electronic transition dipole, characterizing the transition between the initial and final electronic state, does not change with nuclear coordinates of D.
We replace $\varepsilon'$ in this transition dipole moment with its counterpart for a vertical transition, which allows the vibronic transition of D to be written as a product of a purely electronic part and a Franck-Condon factor $|\langle \nu_\mathrm{D} | \nu_{\mathrm{D}^+} \rangle |^2$.
The electronic part is then absorbed in the electronic PI cross section of D, $\sigma^\mathrm{PI}_\mathrm{D}$.\cite{jahr2026_intra}

The ICEC cross section for an atom-molecule system in the Franck-Condon model of ICEC is then given as 
\begin{equation}
    \label{eq:icec_FC}
    \sigma(\varepsilon, \,\nu_\mathrm{D} \to \nu_{\mathrm{D}^+}) 
    = 
    \frac{3 c^4 }{4 \pi} \,
    \frac{
    \sigma^\mathrm{PR}_\mathrm{A}(\varepsilon) \; 
    \sigma^\mathrm{PI}_\mathrm{D}(\omega)
    }{\omega^4 \, R^6} \;|\langle \nu_\mathrm{D} | \nu_{\mathrm{D}^+} \rangle |^2,
\end{equation}
which corresponds to Eq.~(33) in Ref.~\onlinecite{jahr2026_intra} for A being an atom.

The excess energy $\omega$ is transferred to the molecule D, which consequently transitions from a vibrational level $\nu_{\mathrm{D}}$ of the electronic ground state to $\nu_{\mathrm{D}^+}$ of its ionized electronic ground state, while emitting an electron.
The kinetic energy of this electron is given by Eq.~\eqref{eq:omega} with 
the vibronic ionization potential $\mathrm{IP}_{\nu_\mathrm{D} \nu_\mathrm{D^+}}$, defined as the energy difference between the neutral and ionized vibronic state of D,
\begin{equation}
\label{eq:final_E_intra}
    \varepsilon' 
    = \omega - \mathrm{IP}_{\nu_\mathrm{D} \nu_\mathrm{D^+}} - \frac{(Z_\rA-1)}{R}.
\end{equation}

In the current work, we consider the transition to all bound vibrational states of \ce{D+} by including the sum over $\nu_\mathrm{D^+}$ and disregarding dissociative final states,
\begin{equation}
\label{eq:sum_final_states}
    \sigma(\varepsilon, \nu_\mathrm{D}) 
    = 
    \frac{3 c^4 }{4 \pi R^6} 
    \frac{\sigma^\mathrm{PR}_\mathrm{A} \sigma^\mathrm{PI}_\mathrm{D}}{\omega^4}
    \sum_{\nu_{\mathrm{D}^+}}
    |\langle\nu_\mathrm{D}|\nu_{\mathrm{D}^+}\rangle|^2 \,
    H(\varepsilon'),
\end{equation}
where a positive outgoing electron energy is enforced by a Heaviside function $H(\varepsilon')$.
This equation corresponds to Eq.~(34) in Ref.~\onlinecite{jahr2026_intra} for atom-molecule systems when dropping the integral over dissociative final states of \ce{D+}.


\section{Computational Details}
\label{sec:comp-details}

We determined the direct ICEC cross section for all possible combinations with an atomic electron acceptor of the units given in the first column of Tab.~\ref{tab:data_references}, using Eq.~\eqref{eq:icec_electronic} for atom-atom systems and Eq.~\eqref{eq:icec_FC} for atom-molecule systems.
In each of these cases, the ICEC cross section was calculated from threshold to a) $\SI{20}{eV}$ above threshold if the necessary PI information was available or b) for as high incoming electron energies as possible.
A constant step size of the incoming electron energy of $\SI{0.01}{eV}$ was used.
In practice, this is achieved by our in-house software PARFAIT\cite{zenodo}, which handles all possible atom-atom or atom-molecule permutations of acceptor and donor species included in our database and is based on numpy\cite{harris_2020_numpy} and scipy\cite{jones_2001_scipy}.

\begin{table*}[!ht]
    \centering
    \caption{References for the photoionization (or photodetachment) cross section data for the atoms and molecules surveyed in this paper. 
    References to the Franck-Condon factors are included for the molecules.}
    \begin{tabular}{c@{\hspace{10pt}}c@{\hspace{10pt}}c@{\hspace{10pt}}c}
        \toprule
        & Atom or Molecule & Reference & Type \\
        \midrule
        \multirow{7}{*}[-4pt]{\begin{sideways}photoionization\end{sideways}}
        & Kr, Xe &  Samson and Stolte \cite{samson2002_raregases} & experiment \\
        & \ce{F-}, \ce{I-} &  Radojevi\'{c} et al. \cite{radojevic1987_halogen} & theory \\
        & \ce{Cl-}, \ce{Br-} & Robinson and Geltman \cite{robinson1967_ions} & theory, scaled to match experiment\cite{mandl1976_halogen}  \\
        & \ce{H-} & Génévriez and Urbain \cite{genevriez2015_hydride} & experiment \\
        & $Z=1-14, 16, 18, 20, 26$ and their most common cations & TOPbase\cite{topbase1992} & theory \\
        \cmidrule{2-4}
        & LiH & Lundsgaard \cite{lundsgaard1999_LiH} & theory, unresolved \\
        & CO, \ce{H2}, NO, \ce{N2}, \ce{O2} & Leiden database \cite{heays2017_photoionization} & experiment, unresolved \\
        \midrule
        \multirow{3}{*}[-1pt]{\begin{sideways}factor\end{sideways}}
        & CO, NO, \ce{O2} & Wacks \cite{wacks1964_FC} & Franck-Condon, theory \\
        & \ce{H2} & ONeil and Reinhardt \cite{oneil1978_H2} & Franck-Condon, theory \\
        & \ce{N2} & O'Keefe et al. \cite{okeeffe2012_N2} & Franck-Condon, experiment \\
        \bottomrule
    \end{tabular}
    \label{tab:data_references}
\end{table*}

The ICEC channel is open when the transferred energy $\omega$, defined in Eq.~\eqref{eq:omega}, is sufficient to ionize the donor, corresponding to a positive outgoing electron energy $\varepsilon'$ in Eq.~\eqref{eq:final_E_electronic} or \eqref{eq:final_E_intra}, equivalently. 
Therefore, the threshold energy $ \varepsilon_\mathrm{t}$ for ICEC is the smallest value for $\varepsilon$ for which $\varepsilon' \geq 0$. 
To calculate $\varepsilon_\mathrm{t}$, we obtained the
necessary electron affinity of the acceptor A, or equivalently the ionization potential (IP) of \ce{A-}, and the IP of the donor D
from NIST\cite{nist_atomic_spectra}. 
For molecules, we used the adiabatic IP when available and otherwise the recommended value from NIST\cite{NISTWebBook}. 
The energy contributions of the vibrational states were added from the solutions of a Morse potential with parameters obtained from Ref.~\onlinecite{huber_molSpectra}.

For the calculation of the ICEC cross section at a specific incoming electron energy $\varepsilon$, the photorecombination (PR) cross section for the acceptor at $\varepsilon$ and photoionization (PI) cross section for the donor at the corresponding transferred energy $\omega$ are required. 
In practice, it is sufficient to provide PI cross sections since the principle of detailed balance relates the PR cross section with the PI cross section of the reverse process,
\begin{equation}\label{eq:detailed_balance}
    \sigma^\mathrm{PR}_\mathrm{A}(\varepsilon)
    = \frac{\omega^2}{2 \varepsilon c^2}\, \frac{g_{\mathrm{A}^-}}{g_\mathrm{A}} \, \sigma^\mathrm{PI}_{\mathrm{A}^-}(\omega),
\end{equation}
where $g_{\mathrm{A}}$ and $g_{\mathrm{A}^-}$ are the multiplicities of $\mathrm{A}$ and $\mathrm{A}^-$, respectively, and $\omega$ and $\varepsilon$ are connected by energy conservation in Eq.~\eqref{eq:omega}.\cite{jahr2026_intra, sobelman1973}
All PI cross sections were sourced from literature, with the references listed in Tab.~ \ref{tab:data_references}. 
When only figures were available, data was extracted from them using the WebPlotDigitizer \cite{WebPlotDigitizer}. 

For molecules, we applied the Franck-Condon model in Eq.~\eqref{eq:icec_FC} with Franck-Condon factors from sources listed in Table~\ref{tab:data_references},
assuming that the molecule is initially in its ground vibrational states.
As an exception, we applied the purely electronic cross section given in Eq.~\eqref{eq:icec_electronic} for systems containing LiH. 

The Franck-Condon model containing only the bound-bound transition is appropriate when dissociation of the molecule during ionization can be neglected.
The sum over the bound-bound Franck-Condon factors in Eq.~\eqref{eq:sum_final_states} then approaches unity and the ICEC cross section will approach the unresolved value in Eq.~\eqref{eq:icec_electronic}.
The sum over the Franck-Condon factors for \ce{CO}, \ce{H2}, \ce{NO}, \ce{N2}, and \ce{O2} is > 0.98 and dissociation can thus be neglected.
For molecules that predominantly dissociate during ionization, however, the sum over bound Franck-Condon factors is considerably smaller than one, as we have seen for LiH\cite{jahr2026_intra}.
Our model in Ref.\onlinecite{jahr2026_intra} can include dissociation but this would require
a significant repeated effort.
In this case, we have shown \cite{jahr2026_intra} that applying the model in Eq.~\eqref{eq:icec_electronic} with the unresolved molecular PI cross section is better suited for
the calculation of absolute ICEC cross sections in the context of a broad scan over many systems.


The experimental and theoretical PI cross sections from literature were interpolated linearly to obtain PI cross sections for any desired
photon energy. 
For some systems, data points in the low photon energy regime are sparse. 
For simplicity, we assume the PI cross section to decrease linearly with increasing photon energy, whereas in reality the cross section is expected to decrease more rapidly.

We evaluate systems by calculating the ratio between the ICEC cross section and the PR cross section and averaging it over the first $\SI{10}{eV}$ of incoming electron energy.
If the ICEC cross section was available for less than the $\SI{10}{eV}$ interval, the system was not considered for the ranking.
Additionally, we filter out systems which have an average ICEC cross section smaller than $10^{-2}\,\mathrm{Mb}$ over the first $\SI{10}{eV}$ of incoming electron energy. 
All calculations were carried out at a distance between the centers of mass of $R = \SI{7}{\angstrom}$. 

We benchmarked our implementation against existing asymptotic calculations for \ce{Ne+ Xe} in Ref.~\onlinecite{gokhberg2010}, reproducing their results.

Finally, it needs to be noted that we did not analyze whether all reported systems can be realized in experiment.

\section{Results}
\label{sec:results}

\subsection{Effects on the ICEC cross section}

We will first discuss a few recurring effects identified in the ICEC cross sections.
These include the energy dependence of the ICEC cross section, the effect of electronic channel openings and vibrational structure, as well as Cooper minima displayed in the photoionization (PI) cross sections of some atoms.

\subsubsection{Energy dependence}\label{sec:results_omega}

\begin{figure}[!ht]
    \centering
    \includegraphics[width=\linewidth]{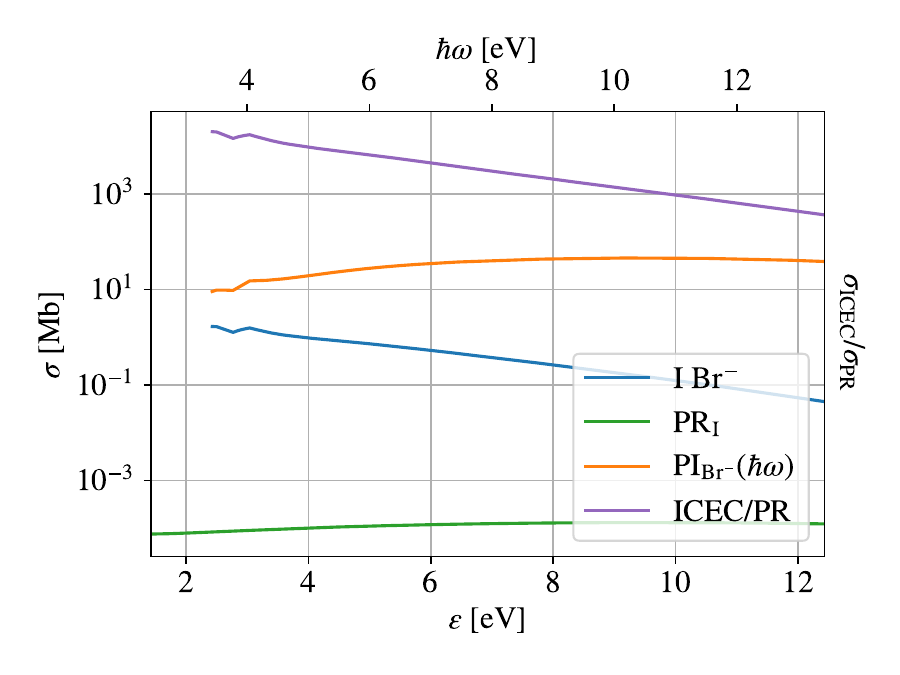}
    \caption{
        ICEC cross section of \ce{I Br-} (blue) against incoming electron energies. Also shown are the photorecombination cross section of I (green), photoionization cross section of \ce{Br-} (orange), and the ICEC/PR ratio (purple).
        ICEC decreases with increasing incoming electron energies.
        \figshareAtoms
    }
    \label{fig:hbaromega_dependency}
\end{figure}

As per the asymptotic model in Eq.~\eqref{eq:icec_electronic} the ICEC cross section decreases with $\omega^{-4}$ and thus with $\varepsilon^{-4}$.
To illustrate this phenomenon, we present the results for the ICEC process \ce{e- + I + Br- \to I- + Br + e-} in Fig.~\ref{fig:hbaromega_dependency}. 

We see this exponential decrease with energy in the ICEC cross section depicted in blue in Fig.~\ref{fig:hbaromega_dependency}.
If the photorecombination (PR) and PI cross sections of the participating units display energy dependencies, these will be reflected in the ICEC cross section additionally to the $\omega^{-4}$ term according to Eq.~\eqref{eq:icec_electronic}. 
Since the PR cross section of \ce{I} (green) and the PI cross section of \ce{Br-} (orange) are nearly flat, the ICEC cross section for \ce{I Br-} decreases almost purely with $\omega^{-4}$.

The ratio between ICEC and PR (purple) is likewise decreasing with $\omega^{-4}$.
In the ratio, only the energy dependence of the PI cross section of the electron donor, here \ce{Br-}, contributes, because the PR cross section is canceled out.


\subsubsection{Electronic channel openings}\label{sec:results_electronic}

Electronic channel openings for the electron donor occur when $\omega$ becomes sufficiently large to ionize \ce{D} into an electronic state which was previously energetically inaccessible. 
If the threshold energy $\varepsilon_\mathrm{t}$ of ICEC is larger than zero, the first observable channel opening corresponds to the onset of the ICEC process itself.
For $\varepsilon_\mathrm{t}=0$, the ICEC channel is always open and subsequent channel openings arise from  electronically excited states of \ce{D+}. 
Each newly accessible electronic state is expected to increase the ICEC cross section by providing additional final state channels.
In our model, these final states are not explicitly calculated, but can be implicitly included via the experimental and theoretical
PI cross sections.

\begin{figure}[t]
    \centering
    \includegraphics[width=0.95\linewidth]{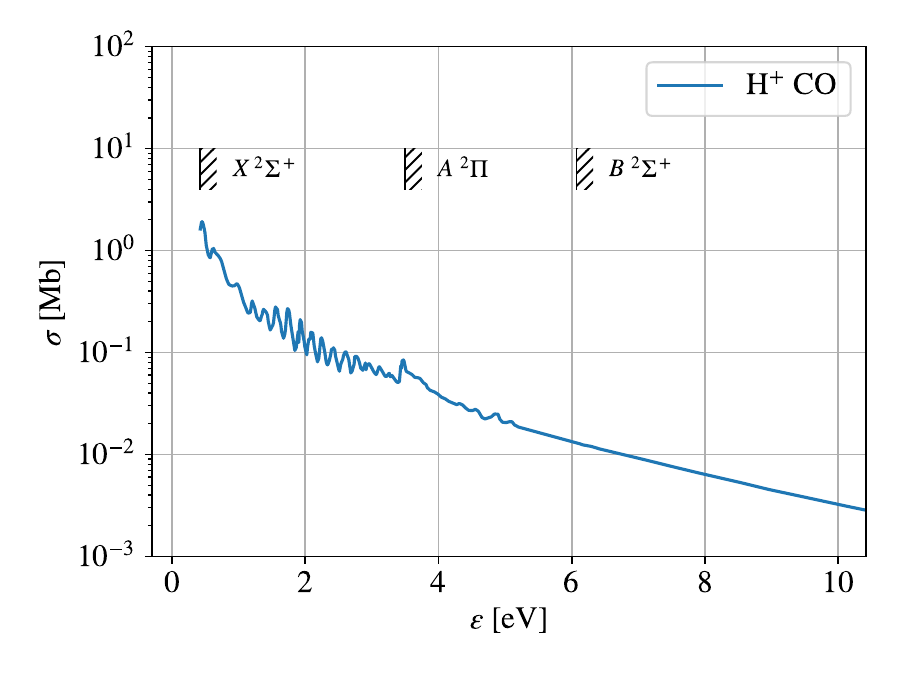}
    \caption{
        ICEC cross section of \ce{H+ CO} against incoming electron energies. 
        Electronic channels are assigned according to Refs.~\onlinecite{heays2017_photoionization} and \onlinecite{chan1993_CO}, however, the effect of $B\,^2\Sigma^+$ is not visible here.
        \figshareMolecules
    }
    \label{fig:electronic_channels}
\end{figure}

As an example, we present the \ce{H+ CO} system in Fig.~\ref{fig:electronic_channels} showing the ICEC cross section with marked channel openings.
At the threshold energy, only the $X\,^2\Sigma^+$ state of \ce{CO+} is accessible.
At $\varepsilon = \SI{3.5}{eV}$, an increase can be observed in the ICEC cross section, where the electronically excited state $A\,^2\Pi$ of \ce{CO+} becomes energetically accessible in the PI cross section.
The electronic channel to the $B\,^2\Sigma^+$ state opens at approximately $\varepsilon = \SI{6}{eV}$, but its impact is hardly visible here.
\cite{heays2017_photoionization, chan1993_CO}

It is important to note that, if the PI cross section of the electron acceptor contains electronic channel openings, converting the PI to the PR cross section gives unphysical results.
In this case, the PI cross section not only describes \ce{A- \to A}, but also \ce{A- \to A^*}.
The reverse process in the PR cross section would then include \ce{A^* \to A-} in addition to the expected \ce{A \to A-} process. 
Fortunately, channel openings in the electron acceptor do not occur for our systems, because 
the atomic PI cross sections in our dataset do not feature such electronic channel openings within a nonrelativistic picture. 
However, if we would consider relativistic effects, the levels of some systems, for example \ce{I^-}, would be impacted by spin-orbit splitting, resulting in two channels with different threshold energies.
Furthermore, we expect such channel openings to occur when considering molecules as electron acceptors.


\subsubsection{Vibrational structure}
\label{sec:results_vibrational}

For atom-molecule systems, vibrational features arise in the ICEC cross section when they are present in the PI cross section of the molecule.
Such features can, for example, arise from autoionizing Rydberg-like states of molecules that lie energetically above the ionized ground state. \cite{holland1993_O2}
The molecule then decays to the ionized ground state by emitting an electron.
During the excitation, the molecule can transition into different vibrational states $\mathrm{v}^\prime$ of the Rydberg state. 
Whenever the photon energy matches a vibronic transition to the Rydberg state, the PI cross section can exhibit a maximum at this energy, leading to a vibrational progression.

\begin{figure}[!ht]
    \centering
    \includegraphics[width=\linewidth]{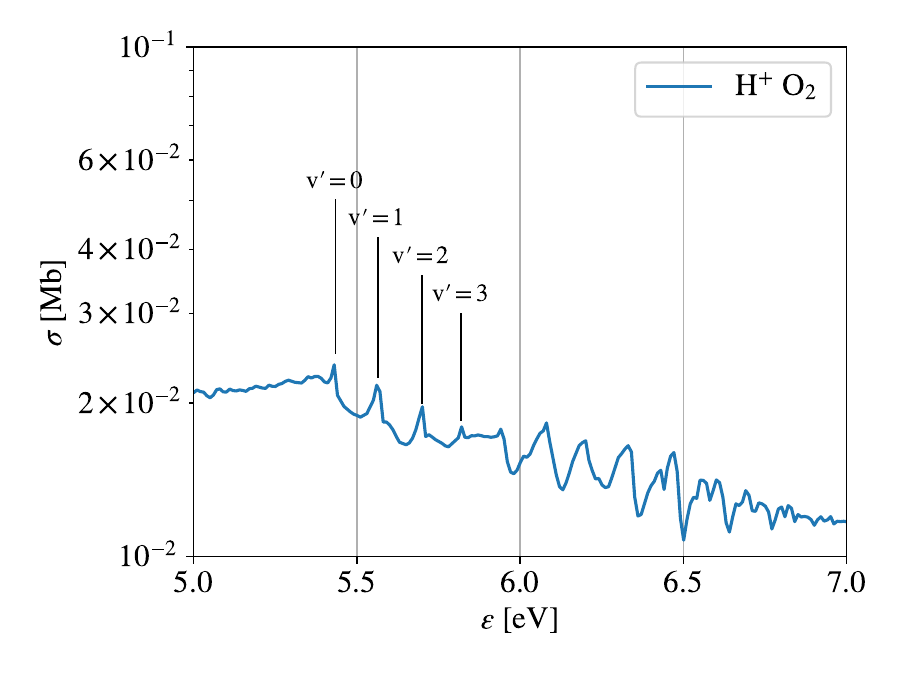}
    \caption{
        ICEC cross section of \ce{H+ O2} against incoming electron energies.
        The annotated peaks correspond to resonant pathways via an auto-ionizing Rydberg state of \ce{O2+}\cite{holland1993_O2, Yoshino_1968}.
        \figshareMolecules
    }
    \label{fig:vibrational_features}
\end{figure}

Even though resonances are not explicitly treated in our theory, their effects appear in the ICEC cross section according to Eq.~\eqref{eq:icec_FC} if they are present in the PI cross section.
This is the case for \ce{H+ O2} presented in Fig.~\ref{fig:vibrational_features}.
Here, an electron of \ce{O2} is excited into the $\mathrm{5p\sigma_u}$ Rydberg state, which subsequently decays to the energetically lower electronic ground state of \ce{O2+}.
Whenever the energy matches a vibronic transition to the Rydberg state, the PI cross section of \ce{O2} -- and subsequently the ICEC cross section -- features a maximum.
In Fig.~\ref{fig:vibrational_features}, the annotated peaks correspond to the vibronic transition from $\nu=0$ of the electronic ground state of \ce{O2} to $\nu^\prime$ of the Rydberg state.
\cite{holland1993_O2, Yoshino_1968}

\subsubsection{Cooper minimum}
\label{sec:results_cooper}
In some systems, we see a sharp minimum in the ICEC cross section which are caused by a Cooper minimum in the PI or PR cross section of one of the participating atoms.
The Cooper minimum is the photon (or electron) energy at which the PI (or PR) cross section experiences a minimum due to a vanishing radial transition matrix element.
This is only possible if the radial part of the emitting (or capturing) orbital has at least one node.
The radial part of the continuum electron wavefunction asymptotically approaches a sine wave, with the frequency depending on the electron energy.
As the shape of the continuum wavefunction changes with energy, the wavefunction can overlap with the radial part of the orbital such that the radial transition matrix element integrates to zero.
The cross section then experiences a minimum at that energy.
\cite{gorczyca2024_sodium}
In general, the PI cross sections from subshells whose radial wave functions are nodeless (such as $1s$, $2p$, $3d$) have a different shape than those from subshells whose radial wave functions have nodes.\cite{cooper1962}

\begin{figure}[!t]
    \centering
    \includegraphics[width=\linewidth]{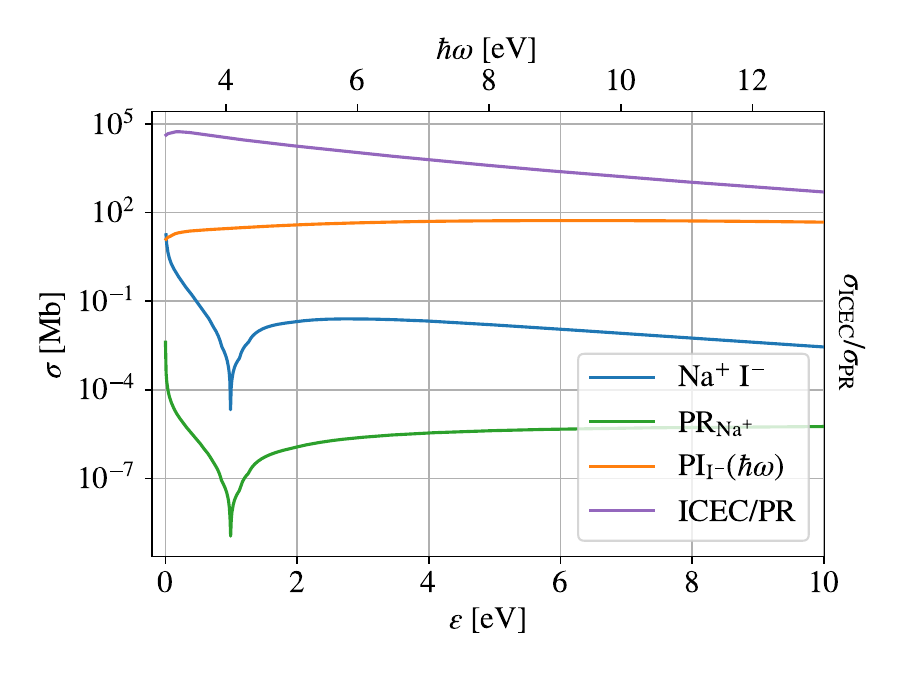}
    \caption{
        ICEC cross section of \ce{Na+ I-} (blue) against incoming electron energies. 
        The Cooper minimum at $\varepsilon = \SI{1}{eV}$ arises from the photorecombination cross section of \ce{Na+} (green).
        Also shown are the photoionization cross section of \ce{I-} (orange) and the ICEC/PR ratio (purple).
        \figshareAtoms
    }
    \label{fig:cooper_minimum}
\end{figure}

We present the ICEC cross section for \ce{Na+ I- \to Na I} as an example in Fig.~\ref{fig:cooper_minimum}.
Here, \ce{Na+} captures into the $3s$ subshell featuring two radial nodes.\cite{gorczyca2024_sodium}
Consequently, both the PR cross section of \ce{Na+} (green) and the ICEC cross section (blue) display a pronounced Cooper minimum at an electron energy of $\varepsilon = \SI{1}{eV}$.


Such Cooper minima in the ICEC cross section might, however, vanish at closer distances between the electron acceptor and donor.
Once the orbitals of the two units overlap, the electron is described by a wavefunction delocalized over both units, reducing the radial symmetry.
However, as long as the asymptotic approximation holds, the two units are treated separately and the Cooper minimum could be visible.

\subsection{Promising candidates for atom-atom systems}
\label{sec:atom-atom_candidates}


The atom-atom systems were ranked according to the criteria explained in Sec.~\ref{sec:comp-details}.
A list of the best 100 atom-atom systems can be found in Tab.~\ref{tab:atom-atom-ranking}.
We will highlight a few promising systems obtained from our ranking.

\begin{table*}[p]
\centering
\caption{
    100 most promising systems for ICEC consisting of two atoms.
    Only systems with an averaged absolute ICEC cross sections $\overline{\sigma}$ of at least $10^{-2}~\mathrm{Mb}$ are included.
    Ranked according to the averaged ratio between the ICEC and the photorecombination cross section of the electron acceptor $\overline{\sigma/\sigma_\mathrm{PR}}$.
    Dataset \textit{anions} is from Refs.~\onlinecite{robinson1967_ions, mandl1976_halogen, radojevic1987_halogen}, $\mathit{noble gases}$ is from Ref.~\onlinecite{samson2002_raregases}, \textit{TOPbase} is from Ref.~\onlinecite{topbase1992}.
    For a detailed list of references see Tab.~\ref{tab:data_references}.
} 
\begin{tabular}{
        cll@{\hspace{5pt}}l@{\hspace{7pt}}l@{\hspace{7pt}}ll@{\hspace{10pt}}|
        @{\hspace{5pt}}cll@{\hspace{5pt}}l@{\hspace{7pt}}l@{\hspace{7pt}}ll
    }
    \toprule
    Rank & \multicolumn{2}{l}{System} & $\overline{\sigma/\sigma_\mathrm{PR}}$ & $\overline{\sigma}~[\mathrm{Mb}]$ & \multicolumn{2}{l}{Datasets (A, D)} 
    & Rank & \multicolumn{2}{l}{System} & $\overline{\sigma/\sigma_\mathrm{PR}}$ & $\overline{\sigma}~[\mathrm{Mb}]$ & \multicolumn{2}{l}{Datasets (A, D)} \\
    \midrule

1 & $\mathrm{F}$ & $\mathrm{I}^-$ & $1.025 \cdot 10^{4}$ & $1.668 \cdot 10^{-1}$ & anions & anions & 51 & $\mathrm{Mg}^+$ & $\mathrm{F}^-$ & $4.133 \cdot 10^{2}$ & $2.099 \cdot 10^{-1}$ & TOPbase & anions \\
2 & $\mathrm{Cl}$ & $\mathrm{I}^-$ & $1.025 \cdot 10^{4}$ & $6.176 \cdot 10^{-1}$ & anions & anions & 52 & $\mathrm{Si}^+$ & $\mathrm{F}^-$ & $3.326 \cdot 10^{2}$ & $1.475 \cdot 10^{1}$ & TOPbase & anions \\
3 & $\mathrm{Br}$ & $\mathrm{I}^-$ & $1.025 \cdot 10^{4}$ & $7.483 \cdot 10^{-1}$ & anions & anions & 53 & $\mathrm{B}^+$ & $\mathrm{F}^-$ & $3.132 \cdot 10^{2}$ & $1.358 \cdot 10^{1}$ & TOPbase & anions \\
4 & $\mathrm{I}$ & $\mathrm{I}^-$ & $1.024 \cdot 10^{4}$ & $1.001 \cdot 10^{0}$ & anions & anions & 54 & $\mathrm{Be}^+$ & $\mathrm{Si}$ & $2.838 \cdot 10^{2}$ & $1.457 \cdot 10^{-1}$ & TOPbase & TOPbase \\
5 & $\mathrm{Na}^+$ & $\mathrm{I}^-$ & $1.024 \cdot 10^{4}$ & $7.866 \cdot 10^{-2}$ & TOPbase & anions & 55 & $\mathrm{I}$ & $\mathrm{S}$ & $2.820 \cdot 10^{2}$ & $3.534 \cdot 10^{-2}$ & anions & TOPbase \\
6 & $\mathrm{Li}^+$ & $\mathrm{I}^-$ & $9.013 \cdot 10^{3}$ & $7.291 \cdot 10^{0}$ & TOPbase & anions & 56 & $\mathrm{Al}^+$ & $\mathrm{S}$ & $2.820 \cdot 10^{2}$ & $8.620 \cdot 10^{-2}$ & TOPbase & TOPbase \\
7 & $\mathrm{Al}^+$ & $\mathrm{I}^-$ & $6.367 \cdot 10^{3}$ & $6.387 \cdot 10^{2}$ & TOPbase & anions & 57 & $\mathrm{B}^+$ & $\mathrm{S}$ & $2.820 \cdot 10^{2}$ & $4.854 \cdot 10^{-1}$ & TOPbase & TOPbase \\
8 & $\mathrm{Ca}^+$ & $\mathrm{I}^-$ & $5.932 \cdot 10^{3}$ & $1.401 \cdot 10^{0}$ & TOPbase & anions & 58 & $\mathrm{Be}^+$ & $\mathrm{S}$ & $2.818 \cdot 10^{2}$ & $1.037 \cdot 10^{-2}$ & TOPbase & TOPbase \\
9 & $\mathrm{F}$ & $\mathrm{Br}^-$ & $4.618 \cdot 10^{3}$ & $8.171 \cdot 10^{-2}$ & anions & anions & 59 & $\mathrm{Si}^+$ & $\mathrm{S}$ & $2.818 \cdot 10^{2}$ & $1.965 \cdot 10^{-1}$ & TOPbase & TOPbase \\
10 & $\mathrm{Li}^+$ & $\mathrm{Br}^-$ & $4.618 \cdot 10^{3}$ & $7.020 \cdot 10^{-1}$ & TOPbase & anions & 60 & $\mathrm{Li}^+$ & $\mathrm{S}$ & $2.818 \cdot 10^{2}$ & $1.487 \cdot 10^{-2}$ & TOPbase & TOPbase \\
11 & $\mathrm{Cl}$ & $\mathrm{Br}^-$ & $4.618 \cdot 10^{3}$ & $3.208 \cdot 10^{-1}$ & anions & anions & 61 & $\mathrm{S}^+$ & $\mathrm{S}$ & $2.818 \cdot 10^{2}$ & $2.664 \cdot 10^{0}$ & TOPbase & TOPbase \\
12 & $\mathrm{Br}$ & $\mathrm{Br}^-$ & $4.618 \cdot 10^{3}$ & $3.788 \cdot 10^{-1}$ & anions & anions & 62 & $\mathrm{Xe}^+$ & $\mathrm{I}^-$ & $2.511 \cdot 10^{2}$ & $1.903 \cdot 10^{-1}$ & noble gases & anions \\
13 & $\mathrm{I}$ & $\mathrm{Br}^-$ & $4.614 \cdot 10^{3}$ & $4.825 \cdot 10^{-1}$ & anions & anions & 63 & $\mathrm{C}^+$ & $\mathrm{S}$ & $2.375 \cdot 10^{2}$ & $5.552 \cdot 10^{-1}$ & TOPbase & TOPbase \\
14 & $\mathrm{Al}^+$ & $\mathrm{Br}^-$ & $3.788 \cdot 10^{3}$ & $2.989 \cdot 10^{2}$ & TOPbase & anions & 64 & $\mathrm{H}^+$ & $\mathrm{I}^-$ & $2.366 \cdot 10^{2}$ & $3.085 \cdot 10^{0}$ & TOPbase & anions \\
15 & $\mathrm{Ca}^+$ & $\mathrm{Br}^-$ & $3.576 \cdot 10^{3}$ & $7.205 \cdot 10^{-1}$ & TOPbase & anions & 65 & $\mathrm{I}$ & $\mathrm{B}$ & $2.350 \cdot 10^{2}$ & $2.832 \cdot 10^{-2}$ & anions & TOPbase \\
16 & $\mathrm{F}$ & $\mathrm{Cl}^-$ & $3.208 \cdot 10^{3}$ & $5.895 \cdot 10^{-2}$ & anions & anions & 66 & $\mathrm{O}^+$ & $\mathrm{I}^-$ & $2.349 \cdot 10^{2}$ & $2.280 \cdot 10^{0}$ & TOPbase & anions \\
17 & $\mathrm{Li}^+$ & $\mathrm{Cl}^-$ & $3.208 \cdot 10^{3}$ & $3.646 \cdot 10^{-1}$ & TOPbase & anions & 67 & $\mathrm{Mg}^+$ & $\mathrm{Al}$ & $2.309 \cdot 10^{2}$ & $1.761 \cdot 10^{-1}$ & TOPbase & TOPbase \\
18 & $\mathrm{Cl}$ & $\mathrm{Cl}^-$ & $3.208 \cdot 10^{3}$ & $2.347 \cdot 10^{-1}$ & anions & anions & 68 & $\mathrm{I}$ & $\mathrm{Ca}$ & $2.125 \cdot 10^{2}$ & $2.060 \cdot 10^{-2}$ & anions & TOPbase \\
19 & $\mathrm{I}$ & $\mathrm{Cl}^-$ & $3.205 \cdot 10^{3}$ & $3.428 \cdot 10^{-1}$ & anions & anions & 69 & $\mathrm{Cl}$ & $\mathrm{Ca}$ & $2.125 \cdot 10^{2}$ & $1.206 \cdot 10^{-2}$ & anions & TOPbase \\
20 & $\mathrm{Al}^+$ & $\mathrm{Cl}^-$ & $2.882 \cdot 10^{3}$ & $2.536 \cdot 10^{2}$ & TOPbase & anions & 70 & $\mathrm{Be}^+$ & $\mathrm{F}^-$ & $2.124 \cdot 10^{2}$ & $5.857 \cdot 10^{-2}$ & TOPbase & anions \\
21 & $\mathrm{Ca}^+$ & $\mathrm{Cl}^-$ & $2.711 \cdot 10^{3}$ & $5.434 \cdot 10^{-1}$ & TOPbase & anions & 71 & $\mathrm{Si}^+$ & $\mathrm{B}$ & $1.974 \cdot 10^{2}$ & $1.586 \cdot 10^{0}$ & TOPbase & TOPbase \\
22 & $\mathrm{Mg}^+$ & $\mathrm{I}^-$ & $2.708 \cdot 10^{3}$ & $1.465 \cdot 10^{0}$ & TOPbase & anions & 72 & $\mathrm{Li}^+$ & $\mathrm{B}$ & $1.974 \cdot 10^{2}$ & $1.332 \cdot 10^{-2}$ & TOPbase & TOPbase \\
23 & $\mathrm{Si}^+$ & $\mathrm{I}^-$ & $2.133 \cdot 10^{3}$ & $1.033 \cdot 10^{2}$ & TOPbase & anions & 73 & $\mathrm{Al}^+$ & $\mathrm{B}$ & $1.973 \cdot 10^{2}$ & $3.726 \cdot 10^{-1}$ & TOPbase & TOPbase \\
24 & $\mathrm{B}^+$ & $\mathrm{I}^-$ & $1.993 \cdot 10^{3}$ & $9.525 \cdot 10^{1}$ & TOPbase & anions & 74 & $\mathrm{B}^+$ & $\mathrm{B}$ & $1.973 \cdot 10^{2}$ & $1.080 \cdot 10^{1}$ & TOPbase & TOPbase \\
25 & $\mathrm{Mg}^+$ & $\mathrm{Br}^-$ & $1.936 \cdot 10^{3}$ & $8.516 \cdot 10^{-1}$ & TOPbase & anions & 75 & $\mathrm{Xe}^+$ & $\mathrm{Xe}$ & $1.798 \cdot 10^{2}$ & $1.334 \cdot 10^{-1}$ & noble gases & noble gases \\
26 & $\mathrm{I}$ & $\mathrm{Al}$ & $1.536 \cdot 10^{3}$ & $1.499 \cdot 10^{-1}$ & anions & TOPbase & 76 & $\mathrm{Si}^+$ & $\mathrm{Xe}$ & $1.798 \cdot 10^{2}$ & $6.070 \cdot 10^{-2}$ & TOPbase & noble gases \\
27 & $\mathrm{F}$ & $\mathrm{Al}$ & $1.534 \cdot 10^{3}$ & $2.392 \cdot 10^{-2}$ & anions & TOPbase & 77 & $\mathrm{B}^+$ & $\mathrm{Xe}$ & $1.798 \cdot 10^{2}$ & $2.172 \cdot 10^{-1}$ & TOPbase & noble gases \\
28 & $\mathrm{Cl}$ & $\mathrm{Al}$ & $1.534 \cdot 10^{3}$ & $8.835 \cdot 10^{-2}$ & anions & TOPbase & 78 & $\mathrm{S}^+$ & $\mathrm{Xe}$ & $1.798 \cdot 10^{2}$ & $8.934 \cdot 10^{-1}$ & TOPbase & noble gases \\
29 & $\mathrm{Mg}^+$ & $\mathrm{Cl}^-$ & $1.512 \cdot 10^{3}$ & $5.969 \cdot 10^{-1}$ & TOPbase & anions & 79 & $\mathrm{C}^+$ & $\mathrm{Xe}$ & $1.798 \cdot 10^{2}$ & $2.195 \cdot 10^{-1}$ & TOPbase & noble gases \\
30 & $\mathrm{Li}^+$ & $\mathrm{Al}$ & $1.307 \cdot 10^{3}$ & $1.646 \cdot 10^{-1}$ & TOPbase & TOPbase & 80 & $\mathrm{I}$ & $\mathrm{Xe}$ & $1.796 \cdot 10^{2}$ & $2.124 \cdot 10^{-2}$ & anions & noble gases \\
31 & $\mathrm{Al}^+$ & $\mathrm{Al}$ & $1.306 \cdot 10^{3}$ & $2.448 \cdot 10^{2}$ & TOPbase & TOPbase & 81 & $\mathrm{Al}^+$ & $\mathrm{Xe}$ & $1.796 \cdot 10^{2}$ & $1.896 \cdot 10^{-2}$ & TOPbase & noble gases \\
32 & $\mathrm{Be}^+$ & $\mathrm{I}^-$ & $1.261 \cdot 10^{3}$ & $4.035 \cdot 10^{-1}$ & TOPbase & anions & 82 & $\mathrm{Si}^+$ & $\mathrm{Al}$ & $1.734 \cdot 10^{2}$ & $7.940 \cdot 10^{0}$ & TOPbase & TOPbase \\
33 & $\mathrm{Ca}^+$ & $\mathrm{Al}$ & $1.143 \cdot 10^{3}$ & $4.928 \cdot 10^{-1}$ & TOPbase & TOPbase & 83 & $\mathrm{H}^+$ & $\mathrm{Xe}$ & $1.702 \cdot 10^{2}$ & $2.059 \cdot 10^{0}$ & TOPbase & noble gases \\
34 & $\mathrm{I}$ & $\mathrm{F}^-$ & $9.868 \cdot 10^{2}$ & $1.043 \cdot 10^{-1}$ & anions & anions & 84 & $\mathrm{O}^+$ & $\mathrm{Xe}$ & $1.691 \cdot 10^{2}$ & $1.536 \cdot 10^{0}$ & TOPbase & noble gases \\
35 & $\mathrm{F}$ & $\mathrm{F}^-$ & $9.859 \cdot 10^{2}$ & $1.755 \cdot 10^{-2}$ & anions & anions & 85 & $\mathrm{N}^+$ & $\mathrm{I}^-$ & $1.676 \cdot 10^{2}$ & $1.927 \cdot 10^{-1}$ & TOPbase & anions \\
36 & $\mathrm{Li}^+$ & $\mathrm{F}^-$ & $9.859 \cdot 10^{2}$ & $1.124 \cdot 10^{-1}$ & TOPbase & anions & 86 & $\mathrm{B}^+$ & $\mathrm{Al}$ & $1.659 \cdot 10^{2}$ & $6.316 \cdot 10^{0}$ & TOPbase & TOPbase \\
37 & $\mathrm{Cl}$ & $\mathrm{F}^-$ & $9.859 \cdot 10^{2}$ & $6.945 \cdot 10^{-2}$ & anions & anions & 87 & $\mathrm{Kr}^+$ & $\mathrm{I}^-$ & $1.619 \cdot 10^{2}$ & $1.512 \cdot 10^{-1}$ & noble gases & anions \\
38 & $\mathrm{Al}^+$ & $\mathrm{F}^-$ & $9.093 \cdot 10^{2}$ & $6.536 \cdot 10^{1}$ & TOPbase & anions & 88 & $\mathrm{Xe}^+$ & $\mathrm{S}$ & $1.521 \cdot 10^{2}$ & $1.103 \cdot 10^{-1}$ & noble gases & TOPbase \\
39 & $\mathrm{Ca}^+$ & $\mathrm{F}^-$ & $8.645 \cdot 10^{2}$ & $1.868 \cdot 10^{-1}$ & TOPbase & anions & 89 & $\mathrm{Al}^+$ & $\mathrm{Ca}$ & $1.465 \cdot 10^{2}$ & $2.577 \cdot 10^{0}$ & TOPbase & TOPbase \\
40 & $\mathrm{I}$ & $\mathrm{Si}$ & $7.370 \cdot 10^{2}$ & $8.795 \cdot 10^{-2}$ & anions & TOPbase & 90 & $\mathrm{Ca}^+$ & $\mathrm{Ca}$ & $1.464 \cdot 10^{2}$ & $1.141 \cdot 10^{-1}$ & TOPbase & TOPbase \\
41 & $\mathrm{F}$ & $\mathrm{Si}$ & $7.363 \cdot 10^{2}$ & $1.526 \cdot 10^{-2}$ & anions & TOPbase & 91 & $\mathrm{Li}^+$ & $\mathrm{Ca}$ & $1.464 \cdot 10^{2}$ & $1.667 \cdot 10^{-2}$ & TOPbase & TOPbase \\
42 & $\mathrm{S}^+$ & $\mathrm{I}^-$ & $7.349 \cdot 10^{2}$ & $5.528 \cdot 10^{0}$ & TOPbase & anions & 92 & $\mathrm{H}^+$ & $\mathrm{S}$ & $1.448 \cdot 10^{2}$ & $1.192 \cdot 10^{0}$ & TOPbase & TOPbase \\
43 & $\mathrm{Mg}^+$ & $\mathrm{Si}$ & $6.844 \cdot 10^{2}$ & $1.396 \cdot 10^{-2}$ & TOPbase & TOPbase & 93 & $\mathrm{O}^+$ & $\mathrm{S}$ & $1.442 \cdot 10^{2}$ & $1.678 \cdot 10^{0}$ & TOPbase & TOPbase \\
44 & $\mathrm{Al}^+$ & $\mathrm{Si}$ & $6.844 \cdot 10^{2}$ & $1.461 \cdot 10^{0}$ & TOPbase & TOPbase & 94 & $\mathrm{Mg}^{2+}$ & $\mathrm{I}^-$ & $1.421 \cdot 10^{2}$ & $7.288 \cdot 10^{-2}$ & TOPbase & anions \\
45 & $\mathrm{Si}^+$ & $\mathrm{Si}$ & $6.837 \cdot 10^{2}$ & $4.899 \cdot 10^{1}$ & TOPbase & TOPbase & 95 & $\mathrm{S}^+$ & $\mathrm{F}^-$ & $1.388 \cdot 10^{2}$ & $9.785 \cdot 10^{-1}$ & TOPbase & anions \\
46 & $\mathrm{Ca}^+$ & $\mathrm{Si}$ & $6.837 \cdot 10^{2}$ & $1.340 \cdot 10^{-2}$ & TOPbase & TOPbase & 96 & $\mathrm{N}^+$ & $\mathrm{Xe}$ & $1.235 \cdot 10^{2}$ & $1.393 \cdot 10^{-1}$ & TOPbase & noble gases \\
47 & $\mathrm{Li}^+$ & $\mathrm{Si}$ & $6.837 \cdot 10^{2}$ & $4.883 \cdot 10^{-2}$ & TOPbase & TOPbase & 97 & $\mathrm{Kr}^+$ & $\mathrm{Xe}$ & $1.195 \cdot 10^{2}$ & $1.101 \cdot 10^{-1}$ & noble gases & noble gases \\
48 & $\mathrm{B}^+$ & $\mathrm{Si}$ & $6.151 \cdot 10^{2}$ & $4.442 \cdot 10^{1}$ & TOPbase & TOPbase & 98 & $\mathrm{Be}^+$ & $\mathrm{B}$ & $1.154 \cdot 10^{2}$ & $4.212 \cdot 10^{-2}$ & TOPbase & TOPbase \\
49 & $\mathrm{C}^+$ & $\mathrm{I}^-$ & $5.279 \cdot 10^{2}$ & $1.809 \cdot 10^{0}$ & TOPbase & anions & 99 & $\mathrm{Ar}^+$ & $\mathrm{I}^-$ & $1.105 \cdot 10^{2}$ & $8.025 \cdot 10^{-1}$ & TOPbase & anions \\
50 & $\mathrm{Ca}^{2+}$ & $\mathrm{I}^-$ & $4.487 \cdot 10^{2}$ & $2.820 \cdot 10^{-1}$ & TOPbase & anions & 100 & $\mathrm{S}^+$ & $\mathrm{Si}$ & $1.095 \cdot 10^{2}$ & $8.574 \cdot 10^{-1}$ & TOPbase & TOPbase \\

    \bottomrule
\end{tabular}
\label{tab:atom-atom-ranking}
\\[0.5ex]
\end{table*}

\subsubsection{Halogen halides}

\begin{figure}[!b]
    \centering
    \includegraphics[width=0.95\linewidth]{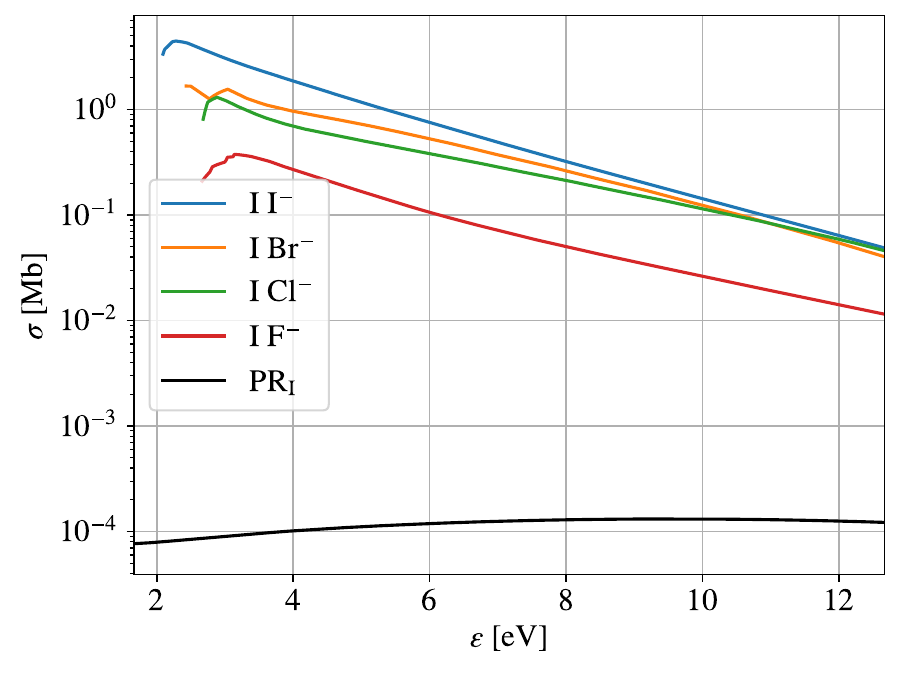}
    \caption{
        ICEC cross sections of selected halogen halides 
        against incoming electron energies;
        \ce{I I-} (blue), 
        \ce{I Br-} (orange), 
        \ce{I Cl-} (green), 
        and \ce{I F-} (red). 
        The photorecombination cross section of \ce{I-} is included in black.
        \figshareAtoms
    }
    \label{fig:halogenides}
\end{figure}

Some of the highest ranking systems resemble halogen halides. 
We show, as representative examples, the ICEC cross sections of \ce{I I-} (blue, ranked fourth in Tab.~\ref{tab:atom-atom-ranking}), \ce{I Br-} (orange, 13th), \ce{I Cl-} (green, 19th) and \ce{I F-} (red, 34th) in Fig.~\ref{fig:halogenides}
and include the PR cross section of I in black for comparison.

All of these systems feature an averaged ratio between the ICEC and PR cross sections on the order of $10^3$.
Moreover, the averaged ICEC cross sections of these systems are between 0.1 and 1\,Mb and therefore lie well within the experimentally accessible range.

While ICEC for \ce{I I-} is allowed at all incoming electron energies, \ce{I Br-}, \ce{I Cl-}, and \ce{I F-} have threshold energies of 0.31, 0.55, and \SI{0.34}{eV}, respectively.
However, the repulsion of the two anions during electron attachment of I decreases the transferred energy by $\SI{2.06}{eV}$ to values below the atomic ionization threshold of the neighbouring halide.
PI cross section data is only available above this atomic threshold, leading to an artificial increase of the ICEC onset by $\SI{2.06}{eV}$.

The \ce{I I-} system (blue) exhibits the largest ICEC cross section and the largest ICEC-to-PR  ratio in this group.
The cross section is dominated by the $\omega^{-4}$ term, as discussed in Sec.~\ref{sec:results_omega}.
However, this effect is dampened due to increases of the PR and PI cross sections of \ce{I} and \ce{I-}, respectively with increasing energy.
As often observed for anions, \ce{I-} exhibits a large PI cross section and a small ionization potential. 
Furthermore, neutral iodine presents a modest electron affinity and a sufficiently large PR cross section.

The ICEC cross sections of \ce{I Br-} (orange) and \ce{I Cl-} (green) behave similarly to the one of \ce{I I-}.
However, both the ICEC cross section and the ICEC-to-PR ratio are reduced by a factor of two compared to the \ce{I I-} system. This can be explained by the low \ce{Br-} and \ce{Cl-}
PI cross sections near threshold.

\ce{I F-} (red) has an ICEC cross section and ICEC/PR ratio further reduced by 75\% compared to the other halogen halide systems.
This is caused by the PI cross section of \ce{F-} being about an order of magnitude smaller than for the other halides, making it a less favourable electron donor.

It should be noted that the real electronic structure of halogen halides are not the neutral-atom-and-ion structure used in this article.
\ce{I_2^-}, in particular, occurs as a molecule in nature.

\subsubsection{Lithium halides}

\begin{figure}[!ht]
    \centering
    \includegraphics[width=0.95\linewidth]{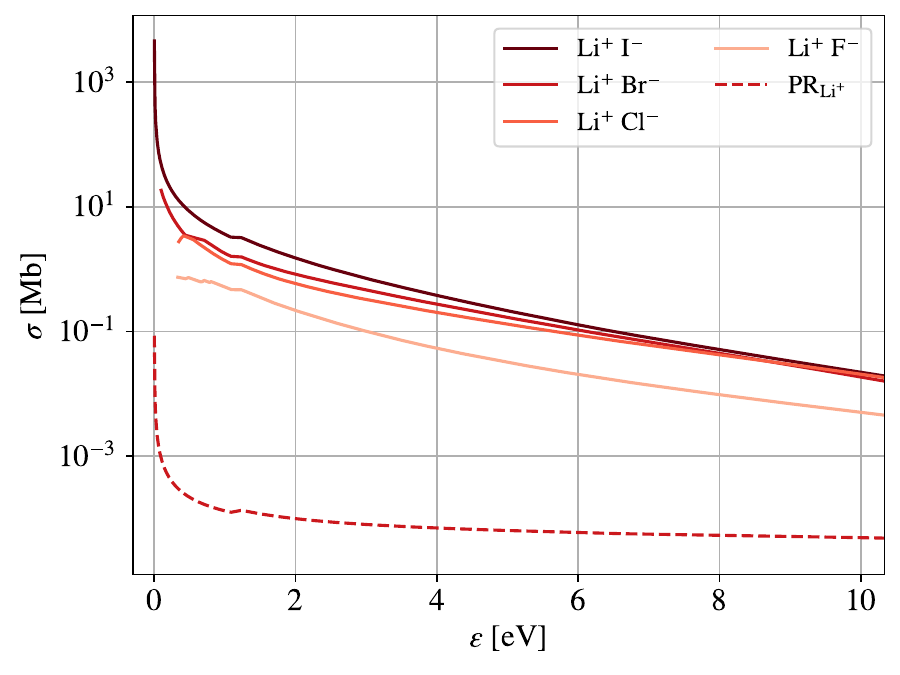}
    \caption{
        ICEC cross sections of lithium halides against incoming electron energies.
        Shading goes from \ce{X+I-} (darkest) to \ce{X+F-} (brightest). 
        Dashed line corresponds to the PR cross sections of the isolated lithium cation.
        \figshareAtoms
    }
    \label{fig:alkali_halides}
\end{figure}

Another class of systems exhibiting promising characteristics are the lithium halides shown in Fig.~\ref{fig:alkali_halides}, with \ce{Li+} acting as the electron acceptor and \ce{F-}, \ce{Cl-}, \ce{Br-}, and \ce{I-} as electron donors.
The PR cross sections of the isolated lithium cation (dashed line) is included for comparison.
We choose to showcase the lithium halides for two reasons: 
First, they rank high in the ICEC-to-PR ratio, meaning ICEC strongly dominates over PR in these systems, occupying the 6th, 10th, 17th and 36rd place, from \ce{I-} to \ce{F-}, in Tab.~\ref{tab:atom-atom-ranking}.
Second, the compounds are chemically stable and water-soluble, which makes them possibly interesting for experiment.

The electron affinity of \ce{Li+} is sufficiently large to ionize \ce{I-} for all incoming electron energies and the ICEC cross section therefore starts at zero incoming electron energy.
For \ce{Br-}, \ce{Cl-}, and \ce{F-}, however, their ionization energies together with the change in attraction is larger than the electron affinity of \ce{Li+} resulting in threshold energies of 0.03, 0.28, and $\SI{0.07}{eV}$, respectively.
Because the PI cross sections of \ce{Br-}, \ce{Cl-}, and \ce{F-} are not available in the literature for the first 0.1-0.3\,eV above threshold, the onsets of the ICEC cross section seen in Fig.~\ref{fig:alkali_halides} differ from the actual threshold energies.

The ICEC cross sections of the lithium halides continuously decrease with increasing incoming electron energies.
The decay is most pronounced for \ce{Li+ I-} close to zero electron energies due to the energy dependence of the PR cross section.


Owing to the large PI cross sections of the halides, the lithium halides exhibit some of the largest ICEC-to-PR ratios in the dataset, despite their ICEC cross sections being only moderate.
Because the transferred energies $\omega$ are small, the ICEC cross section is large enough to dominate over photorecombination over the entire energy range.

\subsection{Promising candidates for atom-molecule systems}

For atom-molecule systems, using the criteria outlined in Sec.~\ref{sec:comp-details}, we discuss promising systems which could be relevant for astro- and atmospheric chemistry.
We present a list of the 96 best performing atom-molecule systems in Tab.~\ref{tab:atom-dimer-ranking}.
Systems containing \ce{LiH} are included for comparison.
A detailed investigation of \ce{H^+ LiH} is presented in our previous publication.\cite{jahr2026_intra}

\afterpage{
\begin{table*}[]
\centering
\caption{96 most promising systems for ICEC consisting of an atom as acceptor and a diatomic molecule as donor. Systems with averaged absolute ICEC cross sections smaller than $10^{-2}~\mathrm{Mb}$ were omitted. Dataset \textit{anions} is from Refs.~\onlinecite{robinson1967_ions, mandl1976_halogen, radojevic1987_halogen}, $\mathit{noble gases}$ is from Ref.~\onlinecite{samson2002_raregases}, \textit{TOPbase} is from Ref.~\onlinecite{topbase1992}, \textit{dimers} is from Ref. \onlinecite{heays2017_photoionization}. For details see Tab.~\ref{tab:data_references}.}
\begin{tabular}{
        cll@{\hspace{5pt}}l@{\hspace{7pt}}l@{\hspace{7pt}}ll@{\hspace{10pt}}|
        @{\hspace{5pt}}cll@{\hspace{5pt}}l@{\hspace{5pt}}l@{\hspace{7pt}}ll
    }
    \toprule
    Rank & \multicolumn{2}{l}{System} & $\overline{\sigma / \sigma_\mathrm{PR}}$ & $\overline{\sigma}~[\mathrm{Mb}]$ & \multicolumn{2}{l}{Datasets (A, D)} 
    & Rank & \multicolumn{2}{l}{System} & $\overline{\sigma / \sigma_\mathrm{PR}}$ & $\overline{\sigma}~[\mathrm{Mb}]$ & \multicolumn{2}{l}{Datasets (A, D)} \\
    \midrule

1 & $\mathrm{I}$ & $\mathrm{LiH}$ & $2.516 \cdot 10^{2}$ & $3.006 \cdot 10^{-2}$ & anions & dimers & 49 & $\mathrm{F}^+$ & $\mathrm{O}_2$ & $3.547 \cdot 10^{1}$ & $8.420 \cdot 10^{-2}$ & TOPbase & dimers \\
2 & $\mathrm{Li}^+$ & $\mathrm{LiH}$ & $2.511 \cdot 10^{2}$ & $1.806 \cdot 10^{-2}$ & TOPbase & dimers & 50 & $\mathrm{F}^+$ & $\mathrm{N}_2$ & $3.530 \cdot 10^{1}$ & $1.402 \cdot 10^{-1}$ & TOPbase & dimers \\
3 & $\mathrm{Al}^+$ & $\mathrm{LiH}$ & $2.511 \cdot 10^{2}$ & $4.995 \cdot 10^{-1}$ & TOPbase & dimers & 51 & $\mathrm{Si}^{2+}$ & $\mathrm{N}_2$ & $3.386 \cdot 10^{1}$ & $1.626 \cdot 10^{-2}$ & TOPbase & dimers \\
4 & $\mathrm{Mg}^+$ & $\mathrm{LiH}$ & $2.509 \cdot 10^{2}$ & $2.107 \cdot 10^{-2}$ & TOPbase & dimers & 52 & $\mathrm{F}^+$ & $\mathrm{CO}$ & $3.198 \cdot 10^{1}$ & $1.482 \cdot 10^{-1}$ & TOPbase & dimers \\
5 & $\mathrm{Si}^+$ & $\mathrm{CO}$ & $7.625 \cdot 10^{1}$ & $1.944 \cdot 10^{-2}$ & TOPbase & dimers & 53 & $\mathrm{F}^+$ & $\mathrm{NO}$ & $3.133 \cdot 10^{1}$ & $5.273 \cdot 10^{-2}$ & TOPbase & dimers \\
6 & $\mathrm{O}^+$ & $\mathrm{CO}$ & $7.625 \cdot 10^{1}$ & $1.350 \cdot 10^{-1}$ & TOPbase & dimers & 54 & $\mathrm{Be}^{2+}$ & $\mathrm{O}_2$ & $3.067 \cdot 10^{1}$ & $5.020 \cdot 10^{-2}$ & TOPbase & dimers \\
7 & $\mathrm{Xe}^+$ & $\mathrm{CO}$ & $7.625 \cdot 10^{1}$ & $4.528 \cdot 10^{-2}$ & noble gases & dimers & 55 & $\mathrm{Be}^{2+}$ & $\mathrm{N}_2$ & $3.014 \cdot 10^{1}$ & $1.093 \cdot 10^{-1}$ & TOPbase & dimers \\
8 & $\mathrm{S}^+$ & $\mathrm{CO}$ & $7.625 \cdot 10^{1}$ & $3.440 \cdot 10^{-1}$ & TOPbase & dimers & 56 & $\mathrm{Be}^{2+}$ & $\mathrm{NO}$ & $2.770 \cdot 10^{1}$ & $3.790 \cdot 10^{-2}$ & TOPbase & dimers \\
9 & $\mathrm{H}^+$ & $\mathrm{CO}$ & $7.625 \cdot 10^{1}$ & $9.510 \cdot 10^{-2}$ & TOPbase & dimers & 57 & $\mathrm{Be}^{2+}$ & $\mathrm{CO}$ & $2.696 \cdot 10^{1}$ & $1.022 \cdot 10^{-1}$ & TOPbase & dimers \\
10 & $\mathrm{C}^+$ & $\mathrm{CO}$ & $7.625 \cdot 10^{1}$ & $7.972 \cdot 10^{-2}$ & TOPbase & dimers & 58 & $\mathrm{C}^+$ & $\mathrm{H}_2$ & $1.556 \cdot 10^{1}$ & $1.258 \cdot 10^{-2}$ & TOPbase & dimers \\
11 & $\mathrm{B}^+$ & $\mathrm{CO}$ & $7.625 \cdot 10^{1}$ & $8.418 \cdot 10^{-2}$ & TOPbase & dimers & 59 & $\mathrm{O}^+$ & $\mathrm{H}_2$ & $1.556 \cdot 10^{1}$ & $2.345 \cdot 10^{-2}$ & TOPbase & dimers \\
12 & $\mathrm{B}^+$ & $\mathrm{NO}$ & $7.534 \cdot 10^{1}$ & $1.419 \cdot 10^{-1}$ & TOPbase & dimers & 60 & $\mathrm{S}^+$ & $\mathrm{H}_2$ & $1.556 \cdot 10^{1}$ & $5.765 \cdot 10^{-2}$ & TOPbase & dimers \\
13 & $\mathrm{Al}^+$ & $\mathrm{NO}$ & $7.533 \cdot 10^{1}$ & $2.554 \cdot 10^{-2}$ & TOPbase & dimers & 61 & $\mathrm{B}^+$ & $\mathrm{H}_2$ & $1.556 \cdot 10^{1}$ & $1.479 \cdot 10^{-2}$ & TOPbase & dimers \\
14 & $\mathrm{Si}^+$ & $\mathrm{NO}$ & $7.532 \cdot 10^{1}$ & $6.627 \cdot 10^{-2}$ & TOPbase & dimers & 62 & $\mathrm{Ar}^+$ & $\mathrm{H}_2$ & $1.515 \cdot 10^{1}$ & $2.477 \cdot 10^{-2}$ & TOPbase & dimers \\
15 & $\mathrm{S}^+$ & $\mathrm{NO}$ & $7.512 \cdot 10^{1}$ & $4.287 \cdot 10^{-1}$ & TOPbase & dimers & 63 & $\mathrm{S}^{2+}$ & $\mathrm{N}_2$ & $1.218 \cdot 10^{1}$ & $7.526 \cdot 10^{-2}$ & TOPbase & dimers \\
16 & $\mathrm{C}^+$ & $\mathrm{NO}$ & $7.201 \cdot 10^{1}$ & $1.675 \cdot 10^{-1}$ & TOPbase & dimers & 64 & $\mathrm{S}^{2+}$ & $\mathrm{NO}$ & $1.168 \cdot 10^{1}$ & $2.972 \cdot 10^{-2}$ & TOPbase & dimers \\
17 & $\mathrm{Ca}^{2+}$ & $\mathrm{NO}$ & $6.841 \cdot 10^{1}$ & $1.033 \cdot 10^{-2}$ & TOPbase & dimers & 65 & $\mathrm{S}^{2+}$ & $\mathrm{O}_2$ & $1.165 \cdot 10^{1}$ & $3.615 \cdot 10^{-2}$ & TOPbase & dimers \\
18 & $\mathrm{N}^+$ & $\mathrm{CO}$ & $6.365 \cdot 10^{1}$ & $6.377 \cdot 10^{-2}$ & TOPbase & dimers & 66 & $\mathrm{S}^{2+}$ & $\mathrm{CO}$ & $1.065 \cdot 10^{1}$ & $6.333 \cdot 10^{-2}$ & TOPbase & dimers \\
19 & $\mathrm{Kr}^+$ & $\mathrm{CO}$ & $6.189 \cdot 10^{1}$ & $5.125 \cdot 10^{-2}$ & noble gases & dimers & 67 & $\mathrm{C}^{2+}$ & $\mathrm{N}_2$ & $1.003 \cdot 10^{1}$ & $5.483 \cdot 10^{-1}$ & TOPbase & dimers \\
20 & $\mathrm{Si}^+$ & $\mathrm{O}_2$ & $5.718 \cdot 10^{1}$ & $1.770 \cdot 10^{-2}$ & TOPbase & dimers & 68 & $\mathrm{F}^+$ & $\mathrm{H}_2$ & $9.898 \cdot 10^{0}$ & $2.140 \cdot 10^{-2}$ & TOPbase & dimers \\
21 & $\mathrm{C}^+$ & $\mathrm{O}_2$ & $5.716 \cdot 10^{1}$ & $7.053 \cdot 10^{-2}$ & TOPbase & dimers & 69 & $\mathrm{He}^+$ & $\mathrm{N}_2$ & $9.671 \cdot 10^{0}$ & $6.753 \cdot 10^{-2}$ & TOPbase & dimers \\
22 & $\mathrm{B}^+$ & $\mathrm{O}_2$ & $5.716 \cdot 10^{1}$ & $6.658 \cdot 10^{-2}$ & TOPbase & dimers & 70 & $\mathrm{C}^{2+}$ & $\mathrm{NO}$ & $9.646 \cdot 10^{0}$ & $2.505 \cdot 10^{-1}$ & TOPbase & dimers \\
23 & $\mathrm{S}^+$ & $\mathrm{O}_2$ & $5.716 \cdot 10^{1}$ & $2.639 \cdot 10^{-1}$ & TOPbase & dimers & 71 & $\mathrm{C}^{2+}$ & $\mathrm{O}_2$ & $9.608 \cdot 10^{0}$ & $2.895 \cdot 10^{-1}$ & TOPbase & dimers \\
24 & $\mathrm{Xe}^+$ & $\mathrm{NO}$ & $5.205 \cdot 10^{1}$ & $3.135 \cdot 10^{-2}$ & noble gases & dimers & 72 & $\mathrm{He}^+$ & $\mathrm{NO}$ & $9.293 \cdot 10^{0}$ & $3.005 \cdot 10^{-2}$ & TOPbase & dimers \\
25 & $\mathrm{Xe}^+$ & $\mathrm{O}_2$ & $5.170 \cdot 10^{1}$ & $2.573 \cdot 10^{-2}$ & noble gases & dimers & 73 & $\mathrm{He}^+$ & $\mathrm{O}_2$ & $9.264 \cdot 10^{0}$ & $3.504 \cdot 10^{-2}$ & TOPbase & dimers \\
26 & $\mathrm{H}^+$ & $\mathrm{O}_2$ & $5.112 \cdot 10^{1}$ & $1.110 \cdot 10^{-1}$ & TOPbase & dimers & 74 & $\mathrm{C}^{2+}$ & $\mathrm{CO}$ & $8.931 \cdot 10^{0}$ & $4.819 \cdot 10^{-1}$ & TOPbase & dimers \\
27 & $\mathrm{O}^+$ & $\mathrm{O}_2$ & $5.101 \cdot 10^{1}$ & $1.555 \cdot 10^{-1}$ & TOPbase & dimers & 75 & $\mathrm{B}^{2+}$ & $\mathrm{N}_2$ & $8.739 \cdot 10^{0}$ & $2.481 \cdot 10^{-2}$ & TOPbase & dimers \\
28 & $\mathrm{H}^+$ & $\mathrm{NO}$ & $5.065 \cdot 10^{1}$ & $1.830 \cdot 10^{-1}$ & TOPbase & dimers & 76 & $\mathrm{He}^+$ & $\mathrm{CO}$ & $8.634 \cdot 10^{0}$ & $5.981 \cdot 10^{-2}$ & TOPbase & dimers \\
29 & $\mathrm{O}^+$ & $\mathrm{NO}$ & $5.045 \cdot 10^{1}$ & $1.837 \cdot 10^{-1}$ & TOPbase & dimers & 77 & $\mathrm{B}^{2+}$ & $\mathrm{O}_2$ & $8.386 \cdot 10^{0}$ & $1.282 \cdot 10^{-2}$ & TOPbase & dimers \\
30 & $\mathrm{N}^+$ & $\mathrm{O}_2$ & $4.921 \cdot 10^{1}$ & $3.398 \cdot 10^{-2}$ & TOPbase & dimers & 78 & $\mathrm{B}^{2+}$ & $\mathrm{NO}$ & $8.369 \cdot 10^{0}$ & $1.126 \cdot 10^{-2}$ & TOPbase & dimers \\
31 & $\mathrm{Kr}^+$ & $\mathrm{O}_2$ & $4.917 \cdot 10^{1}$ & $3.183 \cdot 10^{-2}$ & noble gases & dimers & 79 & $\mathrm{B}^{2+}$ & $\mathrm{CO}$ & $7.867 \cdot 10^{0}$ & $2.191 \cdot 10^{-2}$ & TOPbase & dimers \\
32 & $\mathrm{O}^+$ & $\mathrm{N}_2$ & $4.816 \cdot 10^{1}$ & $7.779 \cdot 10^{-2}$ & TOPbase & dimers & 80 & $\mathrm{Be}^{2+}$ & $\mathrm{H}_2$ & $7.693 \cdot 10^{0}$ & $1.448 \cdot 10^{-2}$ & TOPbase & dimers \\
33 & $\mathrm{B}^+$ & $\mathrm{N}_2$ & $4.816 \cdot 10^{1}$ & $4.410 \cdot 10^{-2}$ & TOPbase & dimers & 81 & $\mathrm{N}^{2+}$ & $\mathrm{O}_2$ & $4.285 \cdot 10^{0}$ & $6.956 \cdot 10^{-2}$ & TOPbase & dimers \\
34 & $\mathrm{N}^+$ & $\mathrm{N}_2$ & $4.816 \cdot 10^{1}$ & $2.719 \cdot 10^{-2}$ & TOPbase & dimers & 82 & $\mathrm{N}^{2+}$ & $\mathrm{N}_2$ & $3.976 \cdot 10^{0}$ & $1.359 \cdot 10^{-1}$ & TOPbase & dimers \\
35 & $\mathrm{Xe}^+$ & $\mathrm{N}_2$ & $4.816 \cdot 10^{1}$ & $1.825 \cdot 10^{-2}$ & noble gases & dimers & 83 & $\mathrm{N}^{2+}$ & $\mathrm{CO}$ & $3.835 \cdot 10^{0}$ & $1.232 \cdot 10^{-1}$ & TOPbase & dimers \\
36 & $\mathrm{H}^+$ & $\mathrm{N}_2$ & $4.816 \cdot 10^{1}$ & $2.042 \cdot 10^{-2}$ & TOPbase & dimers & 84 & $\mathrm{N}^{2+}$ & $\mathrm{NO}$ & $3.799 \cdot 10^{0}$ & $5.729 \cdot 10^{-2}$ & TOPbase & dimers \\
37 & $\mathrm{Kr}^+$ & $\mathrm{N}_2$ & $4.816 \cdot 10^{1}$ & $2.750 \cdot 10^{-2}$ & noble gases & dimers & 85 & $\mathrm{Fe}^{3+}$ & $\mathrm{O}_2$ & $3.717 \cdot 10^{0}$ & $8.228 \cdot 10^{-2}$ & TOPbase & dimers \\
38 & $\mathrm{C}^+$ & $\mathrm{N}_2$ & $4.816 \cdot 10^{1}$ & $3.747 \cdot 10^{-2}$ & TOPbase & dimers & 86 & $\mathrm{Fe}^{3+}$ & $\mathrm{N}_2$ & $3.296 \cdot 10^{0}$ & $1.339 \cdot 10^{-1}$ & TOPbase & dimers \\
39 & $\mathrm{S}^+$ & $\mathrm{N}_2$ & $4.816 \cdot 10^{1}$ & $1.647 \cdot 10^{-1}$ & TOPbase & dimers & 87 & $\mathrm{Fe}^{3+}$ & $\mathrm{CO}$ & $3.264 \cdot 10^{0}$ & $1.262 \cdot 10^{-1}$ & TOPbase & dimers \\
40 & $\mathrm{Ar}^+$ & $\mathrm{CO}$ & $4.772 \cdot 10^{1}$ & $2.155 \cdot 10^{-1}$ & TOPbase & dimers & 88 & $\mathrm{Fe}^{3+}$ & $\mathrm{NO}$ & $3.206 \cdot 10^{0}$ & $6.723 \cdot 10^{-2}$ & TOPbase & dimers \\
41 & $\mathrm{Ar}^+$ & $\mathrm{N}_2$ & $4.720 \cdot 10^{1}$ & $1.638 \cdot 10^{-1}$ & TOPbase & dimers & 89 & $\mathrm{O}^{2+}$ & $\mathrm{O}_2$ & $2.145 \cdot 10^{0}$ & $1.481 \cdot 10^{-2}$ & TOPbase & dimers \\
42 & $\mathrm{Ar}^+$ & $\mathrm{O}_2$ & $4.535 \cdot 10^{1}$ & $7.798 \cdot 10^{-2}$ & TOPbase & dimers & 90 & $\mathrm{O}^{2+}$ & $\mathrm{CO}$ & $1.684 \cdot 10^{0}$ & $2.378 \cdot 10^{-2}$ & TOPbase & dimers \\
43 & $\mathrm{Si}^{2+}$ & $\mathrm{O}_2$ & $4.311 \cdot 10^{1}$ & $1.260 \cdot 10^{-1}$ & TOPbase & dimers & 91 & $\mathrm{O}^{2+}$ & $\mathrm{NO}$ & $1.671 \cdot 10^{0}$ & $1.056 \cdot 10^{-2}$ & TOPbase & dimers \\
44 & $\mathrm{N}^+$ & $\mathrm{NO}$ & $4.286 \cdot 10^{1}$ & $3.225 \cdot 10^{-2}$ & TOPbase & dimers & 92 & $\mathrm{O}^{2+}$ & $\mathrm{N}_2$ & $1.598 \cdot 10^{0}$ & $2.258 \cdot 10^{-2}$ & TOPbase & dimers \\
45 & $\mathrm{Kr}^+$ & $\mathrm{NO}$ & $4.257 \cdot 10^{1}$ & $2.868 \cdot 10^{-2}$ & noble gases & dimers & 93 & $\mathrm{C}^{2+}$ & $\mathrm{H}_2$ & $1.448 \cdot 10^{0}$ & $3.765 \cdot 10^{-2}$ & TOPbase & dimers \\
46 & $\mathrm{Si}^{2+}$ & $\mathrm{CO}$ & $4.219 \cdot 10^{1}$ & $4.048 \cdot 10^{-1}$ & TOPbase & dimers & 94 & $\mathrm{N}^{3+}$ & $\mathrm{O}_2$ & $5.251 \cdot 10^{-1}$ & $1.793 \cdot 10^{-2}$ & TOPbase & dimers \\
47 & $\mathrm{Ar}^+$ & $\mathrm{NO}$ & $3.781 \cdot 10^{1}$ & $6.003 \cdot 10^{-2}$ & TOPbase & dimers & 95 & $\mathrm{N}^{3+}$ & $\mathrm{NO}$ & $4.084 \cdot 10^{-1}$ & $1.234 \cdot 10^{-2}$ & TOPbase & dimers \\
48 & $\mathrm{Si}^{2+}$ & $\mathrm{NO}$ & $3.595 \cdot 10^{1}$ & $8.445 \cdot 10^{-2}$ & TOPbase & dimers & 96 & $\mathrm{N}^{3+}$ & $\mathrm{N}_2$ & $3.892 \cdot 10^{-1}$ & $2.198 \cdot 10^{-2}$ & TOPbase & dimers \\
    \bottomrule
\end{tabular}
\label{tab:atom-dimer-ranking}
\end{table*}

}

\subsubsection{Proton molecule systems}

\begin{figure}[htb]
    \centering
    \includegraphics[width=0.95\linewidth]{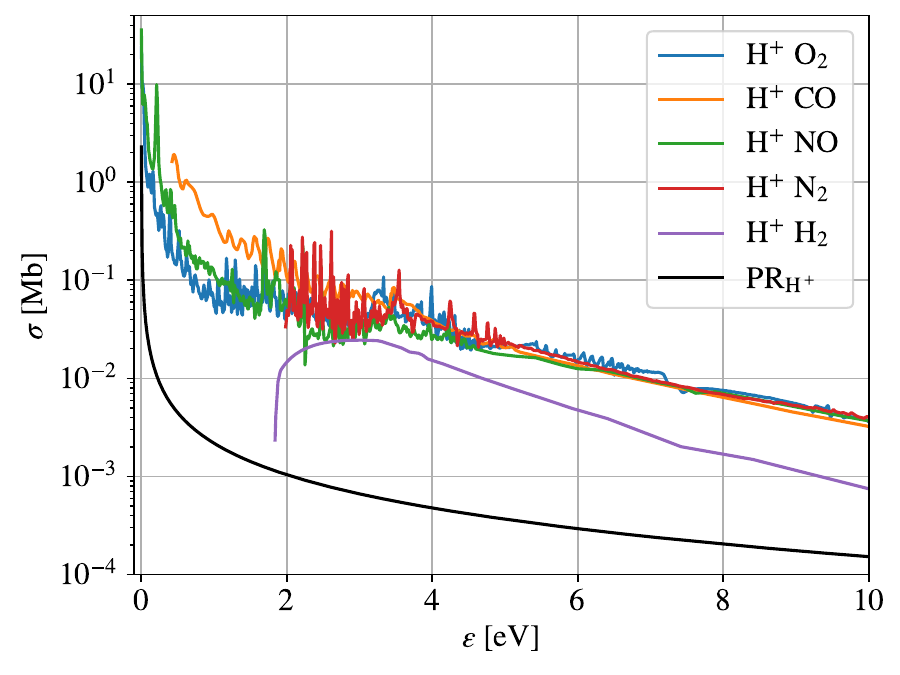}
    \caption{
        ICEC cross sections for atmospherically relevant proton-molecule systems against incoming electron energies up to $\varepsilon = \SI{10}{eV}$.
        \ce{H+ O2} (blue), 
        \ce{H+ CO} (orange), 
        \ce{H+ NO} (green), 
        \ce{H+ N2} (red), 
        and \ce{H+ H2} (purple). 
        The PR cross section of isolated \ce{H+} is shown in black.
        \figshareMolecules
    }
    \label{fig:H+_dimers}
\end{figure}

The first group we present is  a proton in combination with the atmospherically relevant diatomic molecules \ce{O2}, CO, NO, \ce{N2}, and \ce{H2}.
In Fig.~\ref{fig:H+_dimers}, we show the ICEC cross sections for \ce{H+} with \ce{O2} (blue), CO (orange), NO (green), \ce{N2} (red), and \ce{H2} (purple) together with the PR cross section of isolated \ce{H+} (black).

In \ce{H+ O2} and \ce{H+ NO}, ICEC is energetically allowed for all incoming electron energies, while \ce{H+ CO}, \ce{H+ H2}, and \ce{H+ N2} exhibit threshold energies of $\varepsilon_\mathrm{t} = 0.4$, 1.8, and $\SI{1.98}{eV}$, respectively.

\ce{H+ O2}, \ce{H+ CO}, \ce{H+ NO}, and \ce{H+ N2} all exhibit electronic channel openings and vibrational
features inherited from the molecular PI cross sections, as discussed in Secs.~\ref{sec:results_electronic} and
\ref{sec:results_vibrational}. 
Above $\varepsilon = \SI{4}{eV}$, their ICEC cross sections decrease similarly.


The ICEC cross section of \ce{H+ H2} increases sharply right after the ICEC channel opens but remains about half an order of magnitude below the other systems. 
In the displayed energy range, it exhibits no resonant features.

We would like to discuss the possible natural occurrence of some of the mentioned systems. 
The isomer \ce{HCO+} of \ce{H+ CO} plays an important role in astrochemistry.
Theoretical models show that the formyl cation \ce{HCO+} is central in the water depletion of Venus.\cite{Chaffin2024}
In these investigations, the formation of \ce{HCO+} occurs via proton transfer between \ce{H2+} and \ce{CO}\cite{wu_2024}.
While more in depth studies of ICEC in more representative systems like \ce{H2+ CO} are needed, \ce{H+ CO} can be seen as a model for a transition state in that reaction.
Furthermore, the reaction network also includes charge-transfer reactions between \ce{H+} and \ce{O2}, for which ICEC would constitute a competing process.\cite{Chaffin2024,wu_2024}

\ce{H+ H2} is of special interest to the study of hydrogen plasmas and astrochemistry, amongst others in the atmosphere of Jupiter\cite{Tennyson_2019}.
Both \ce{H+ H2} and \ce{H H2+} are dissociation products of \ce{H3+} and engage in charge transfer reactions between each other.
These reactions could be catalyzed by an incoming electron
performing ICEC.

\subsubsection{Oxygen cation molecule systems}

Oxygen cation systems are, just as the proton systems, relevant to atmospheric chemistry and astrochemistry.
\ce{O+} is formed in Earth's upper ionosphere and is the main cation produced at altitudes above 100\,km\cite{pavlov_2014, ferguson_2007}.

\begin{figure}[!hb]
    \centering
    \includegraphics[width=0.95\linewidth]{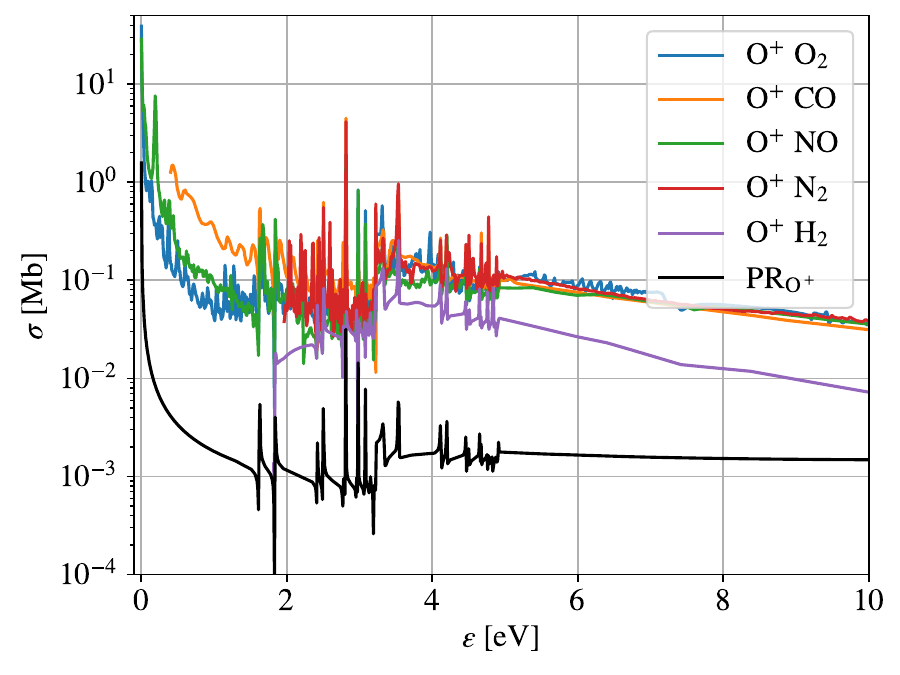}
    \caption{
        ICEC cross sections for atmospherically relevant oxygen-molecule systems against incoming electron energies.
        \ce{O+ O2} (blue), 
        \ce{O+ CO} (orange), 
        \ce{O+ NO} (green), 
        \ce{O+ N2} (red), 
        and \ce{O+H2} (purple).
        The photorecombination cross section of isolated \ce{O+} is shown in black.
        \figshareMolecules
    }
    \label{fig:O+_dimers}
\end{figure}

We therefore present the ICEC cross sections of \ce{O+} with the molecules \ce{O2} (blue), CO (orange), NO (green), \ce{N2} (red), and \ce{H2} (purple) together with the PR cross section of isolated \ce{O+} (black) in Fig.~\ref{fig:O+_dimers}.
The ICEC cross sections of these systems are similar to those of the respective proton systems because
the electron affinities of \ce{O+} ($\SI{13.618}{eV}$) and \ce{H+} ($\SI{13.598}{eV}$) are similar 
and their PR cross sections are in the same order of magnitude.
The main difference in the ICEC cross section are the additional resonance features introduced by the PR cross section of \ce{O+} between $\varepsilon = \SI{1.8}{eV}$ and $\SI{5}{eV}$.
Like the proton molecule systems, the ICEC cross section of the oxygen cation molecule systems is 10 -- 76 times
larger than the photorecombination cross section of the oxygen cation.

ICEC could therefore potentially compete with the charge-transfer reactions
\ce{O+ + O2 \to O + O2+} and \ce{O+ + NO \to O + NO+}, which play a key
role in the ion chemistry of Earth’s upper ionosphere.\cite{st-maurice_1978}

Another important reaction with \ce{O+} is the formation of the nitrosyl cation, \ce{O+ + N2 \to NO+ + N}, in Earth's upper ionosphere. 
As shown above, the calculated ICEC cross section of \ce{O+ N2} suggests that ICEC may compete with the formation of \ce{NO+}.
Considering that \ce{NO+} acts as a control on the removal of ions from the ionosphere \cite{ferguson_2007},
ICEC could thus slow down deionization of the ionosphere.

Furthermore, \ce{O+ H2} could be relevant in Venus' ionosphere, where the proton transfer reaction \ce{O+ + H2 \to OH+ + H} is proposed as the reaction which provides hot hydrogen atoms for various follow-up reactions \cite{kumar1978}.
ICEC could be a competing process (\ce{O+ + H2 \to O + H2+}) or an intermediate process in the proton transfer reaction.


\subsubsection{Predicted spectra}

For atom-atom systems without considering interparticle motion, the ICEC electron spectra for a given incoming electron energy displays a single peak at the outgoing electron energy given in Eq.~\eqref{eq:final_E_electronic}.
For atom-molecule systems, however, the spectra are broadened due to the additional nuclear degree of freedom.
For a given incoming electron energy $\varepsilon$, the outgoing electron energy $\varepsilon'$ is determined by Eq.~\eqref{eq:final_E_intra} for each vibronic transtion of the molecule.
The higher the vibrational final state of the molecule, the lower $\varepsilon'$ is for that transition.
The intensity of the corresponding peak is given by Eq.~\eqref{eq:icec_FC}.

\begin{figure}[!ht]
    \centering
    \includegraphics[width=0.95\linewidth]{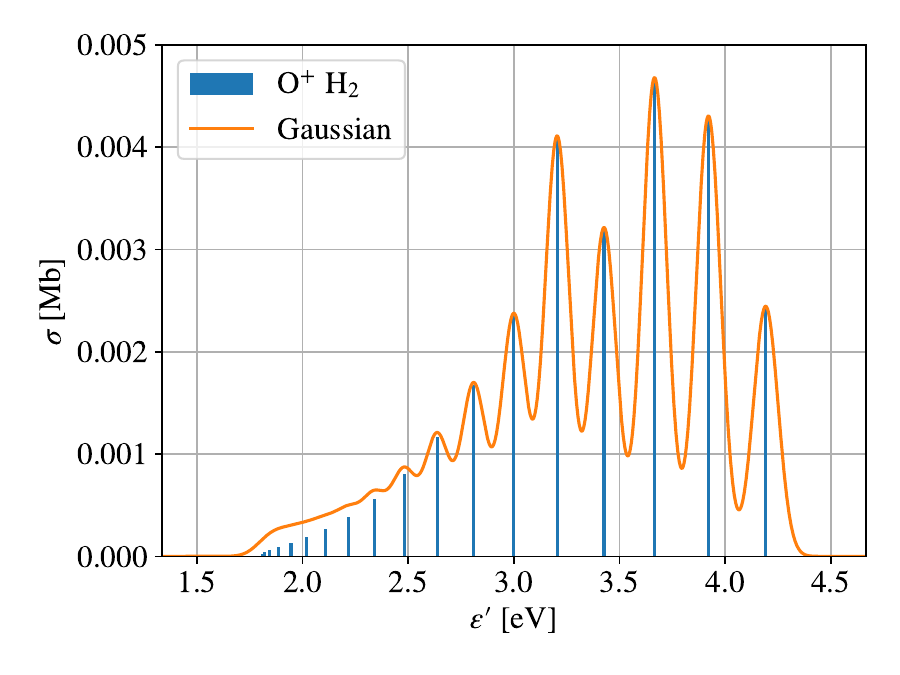}
    \caption{
        ICEC electron spectrum of \ce{O+ H2} (blue) against outgoing electron energies with a fixed incoming electron energy of $\varepsilon = \SI{6}{eV}$.
        The spectrum is folded with a Gaussian envelope with $\mathrm{FWHM} = \SI{0.1}{eV}$ (orange).
        \figshareMolecules
    }
    \label{fig:O+_H2_xs_spectrum}
\end{figure}

Fig.~\ref{fig:O+_H2_xs_spectrum} shows the ICEC electron spectrum of \ce{O+ H2} in blue with a Gaussian envelope ($\mathrm{FWHM} = \SI{0.1}{eV}$) in orange. 
The peak of highest outgoing electron energy corresponds to the adiabatic ionization from $\mathrm{v} = 0$ in \ce{H2} to $\mathrm{v}_+=0$ in \ce{H2+}. 
However, the peak with maximum intensity corresponds to the transition $\mathrm{v} = 0 \to \mathrm{v}_+ = 2$. 
As the final vibrational levels approach the dissociation limit of \ce{H2+}, the spacing between neighbouring bound states decreases and the peaks corresponding to $\mathrm{v}_+\geq 9$ are not distinguishable in the Gaussian envelope.

As discussed in Sec.~\ref{sec:comp-details}, the sum over bound-bound Franck-Condon factors is above 0.98 for all atmospherically relevant dimers. They therefore dominate and we neglect the dissociation of these dimers
in this investigation.
In systems which feature \ce{LiH}, however, dissociation after ionization plays a significant role\cite{jahr2026_intra}. 
Beyond $\varepsilon'=\omega-\rIP_{\nu_\rD, \text{diss}}$, where $\rIP_{\nu_\rD, \text{diss}}$ includes the dissociation energy of \ce{D+}, the final states belong to the dissociative continuum rather than the discrete bound spectrum, giving rise to a continuous spectral profile.

\section{Conclusion}
\label{sec:conclusion}

We systematically evaluated ICEC across 2442 atom–atom and
atom–molecule systems to identify and rank promising candidates for
future experimental and theoretical studies. The systems comprise 23
elements and six diatomic molecules in different charge states. 
To identify candidates that combine large ICEC cross sections with favorable competition against photorecombination, 
we evaluated the ICEC
cross sections and ICEC-to-photorecombination ratios, averaging both
quantities over the first \SI{10}{eV} of incoming electron energy above
threshold.
Motivated by the expected requirements for experimental
observation, we retained systems with an averaged ICEC cross section exceeding
$10^{-2}\,\mathrm{Mb}$ and ranked them according to their averaged
ICEC-to-photorecombination ratio.


While Gokhberg and Cederbaum\cite{gokhberg2010} incorporated the Coulomb attraction and repulsion
between charged units in the threshold energy,
our work is the first to account for these interactions throughout the ICEC process when calculating
ICEC cross sections and electron spectra. If present, these interactions modify the transferred energy $\omega$, thereby affecting
the ICEC cross section through its $\omega^{-4}$ dependence. They also
alter the kinetic energy of the emitted electron and, consequently, the
predicted electron spectrum.

Despite the wide variety of systems considered, several recurring features
emerge in the calculated ICEC cross sections. Their overall energy
dependence is shaped by the $\omega^{-4}$ dependence on the transferred
energy, as well as by the underlying photoionization and
photorecombination cross sections. Additional ICEC channels into
electronically excited states can increase the total cross section,
while vibrational structure, resonances, and Cooper
minima of the isolated units can leave distinct signatures. In the full
interacting system, however, resonant and vibrational features may shift
or change, while Cooper minima may be weakened or absent.

The trends emerging from our broad survey are consistent with the
established characteristics which favour 
ICEC.~\cite{gokhberg2009,bande2023} In particular, small transferred
energies and large photoionization cross sections of the electron donor promote
both a large ICEC cross section and its dominance over
photorecombination.
While the ICEC-to-photorecombination ratio is independent of the photorecombination cross section itself, a larger photorecombination cross section further enhances the absolute ICEC cross section.

Among the atom–atom systems, halogen–halides and lithium
halides emerge as particularly promising candidates for further studies.
Notably, halogen–halide systems, for which ICEC was originally
proposed, exhibit particularly large ICEC cross
sections within our broad survey. Lithium halides further display a strong
dominance of ICEC over photorecombination, making them an interesting
class of systems for further study.

Among the atom--molecule systems, proton- and oxygen-cation systems
with the diatomic molecules \ce{O2}, CO, NO, and \ce{N2} emerge as
particularly promising candidates. Besides exhibiting favourable ICEC
cross sections and ICEC-to-photorecombination ratios, these systems are
relevant to atmospheric and astrochemical environments. ICEC may therefore compete with established charge-transfer
reactions and should be considered in future studies of these
environments.


The resulting database and ranking provide a systematic basis for
selecting promising systems for future theoretical and experimental studies.
More generally, our results demonstrate that ICEC is not
limited to the small number of model systems investigated so far, but
may occur across a much broader range of chemical environments.




\section*{Supplementary Material}

The supplementary material contains the complete set of calculated ICEC
cross sections as a function of the incoming electron energy for all
investigated systems together with the calculated ICEC electron spectra
for all systems with dimer electron donors.

\begin{acknowledgments}
We thank the DFG-ANR for financial support through the QD4ICEC project with grant number FA 1989/1-1.
E. F. furthermore acknowledges funding by LISA$^+$ at the University of Tübingen. 
\end{acknowledgments}

\section*{Author statements}
\subsection*{Conflict of interest}
The authors have no conflicts to disclose.

\subsection*{Author Contributions}
\textbf{Conceptualization}: E.F. (lead), E.M.J.
\textbf{Data curation}: F.L.H.
\textbf{Formal Analysis}: E.M.J. (lead), F.L.H. 
\textbf{Funding}: E.F.
\textbf{Investigation}: E.M.J. (equal), F.L.H. (equal), E.F. (supporting)
\textbf{Methodology}: E.M.J. (equal), E.F. (equal)
\textbf{Software}: E.M.J. (equal), F.L.H. (equal)
\textbf{Supervision}: E.F. (lead), E.M.J. 
\textbf{Validation}: F.L.H.
\textbf{Visualization}: E.M.J. (equal), F.L.H. (equal)
\textbf{Writing -- original draft}: F.L.H. (equal), E.M.J. (equal), E.F. (equal)
\textbf{Writing -- review \& editing}: E.M.J. (equal), E.F. (equal), F.L.H. (contributing)

\section*{Data availability}
The data that support the findings of this study are openly available and cited at the appropriate locations within this paper.

\appendix

\bibliography{refs}

\end{document}